\documentclass[journal,12pt,onecolumn,draftclsnofoot,]{IEEEtran}

\usepackage[cmex10]{amsmath}
\usepackage[utf8]{inputenc}
\usepackage{multirow}
\usepackage{soul}
\usepackage{algorithm}
\usepackage{algpseudocode}
\usepackage{graphicx}
\usepackage{dsfont}
\usepackage{xcolor}
\usepackage{cite}

\usepackage[utf8]{inputenc}
\usepackage{mathtools}
\usepackage[mathscr]{euscript}
\usepackage{amsmath}
\usepackage{amssymb}
\usepackage{amsfonts}
\usepackage{amssymb}
\usepackage{amsmath}
\usepackage{verbatim}
\usepackage[T1]{fontenc}
\usepackage[utf8]{inputenc}
\usepackage{latexsym}
\usepackage{enumerate}
\usepackage{indentfirst}
\usepackage{calligra}
\usepackage{ulem}
\usepackage{graphicx}
\usepackage{array}
\usepackage{float}
\usepackage{bbm}
\usepackage{mleftright}
\usepackage{colortbl}
\makeatletter
\let\old@ps@headings\ps@headings
\let\old@ps@IEEEtitlepagestyle\ps@IEEEtitlepagestyle
\def\psccfooter#1{%
 \def\ps@headings{%
 \old@ps@headings%
 \def\@oddfoot{\strut\hfill#1\hfill\strut}%
 \def\@evenfoot{\strut\hfill#1\hfill\strut}%
 }%
 \def\ps@IEEEtitlepagestyle{%
 \old@ps@IEEEtitlepagestyle%
 \def\@oddfoot{\strut\hfill#1\hfill\strut}%
 \def\@evenfoot{\strut\hfill#1\hfill\strut}%
 }%
 \ps@headings%
}
\makeatother

\usepackage{soul}
\usepackage{multirow}
\usepackage{soul,color}
\usepackage{graphicx}
\usepackage{dsfont}
\usepackage{subcaption}

\usepackage[utf8]{inputenc}
\usepackage{mathtools}
\usepackage[mathscr]{euscript}
\usepackage{amsmath}
\usepackage{amssymb}
\usepackage{amsfonts}
\usepackage{verbatim}
\usepackage[T1]{fontenc}
\usepackage[utf8]{inputenc}

\usepackage{latexsym}
\usepackage{enumerate}
\usepackage{indentfirst}
\usepackage{calligra}
\usepackage{ulem}
\usepackage{multirow}

\usepackage{array}
\usepackage{float}
\usepackage{bbm}

\usepackage{eurosym}

\usepackage{hyperref}

\usepackage{caption}
\usepackage{tikz}
\usepackage{pgfplots}

\pgfplotsset{compat=1.8}
\usepgfplotslibrary{statistics}
\pgfmathdeclarefunction{fpumod}{2}{%
 \pgfmathfloatdivide{#1}{#2}%
 \pgfmathfloatint{\pgfmathresult}%
 \pgfmathfloatmultiply{\pgfmathresult}{#2}%
 \pgfmathfloatsubtract{#1}{\pgfmathresult}%
 \pgfmathfloatifapproxequalrel{\pgfmathresult}{#2}{\def\pgfmathresult{5}}{}%
 }

\usepackage{tabularx}
\usepackage{flushend}
\usepackage{textcomp}
\usepackage{xcolor}
\usepackage{import}

\usetikzlibrary{trees,decorations,shadows}
\tikzset{level 1/.style={sibling angle=45,level distance=4mm}}
\usetikzlibrary{arrows.meta}
\usepackage{forest}
\usetikzlibrary{external}
\let\oldtikzexternalgetnextfilename\tikzexternalgetnextfilename \renewcommand{\tikzexternalgetnextfilename}[1]{\oldtikzexternalgetnextfilename{#1}\expandafter\tikzsetnextfilename\expandafter{#1}}

\usepackage{pgfplotstable}
\usepackage[outline]{contour}
\contourlength{1.2pt}

\pgfplotsset{compat=1.13} 

\usetikzlibrary{spy}
\usetikzlibrary{calc}
\usetikzlibrary{fadings}
\usetikzlibrary{patterns}
\usetikzlibrary{shadows}
\usetikzlibrary{mindmap}
\usetikzlibrary{backgrounds}
\usetikzlibrary{shapes.symbols}
\usetikzlibrary{shapes.multipart}
\usetikzlibrary{shapes.geometric}
\usetikzlibrary{automata,positioning}
\usetikzlibrary{decorations.fractals} 
\usetikzlibrary{decorations.markings}
\usetikzlibrary{decorations.pathreplacing}
\usetikzlibrary{decorations.pathmorphing}
 
\usepackage{bm}

\usepackage{nomencl}
\makenomenclature
\tikzset{edge from parent/.style={segment angle=10,draw}}

\tikzset{
 my rounded corners/.append style={rounded corners=2pt},
}

\def\BibTeX{{\rm B\kern-.05em{\sc i\kern-.025em b}\kern-.08em
 T\kern-.1667em\lower.7ex\hbox{E}\kern-.125emX}}

\renewcommand{\nomgroup}[1]{%
 \ifthenelse{\equal{#1}{O}}{\item[\textit{Operators}]}{%
 \ifthenelse{\equal{#1}{I}}{\item[\textit{Indices}]}{%
 \ifthenelse{\equal{#1}{A}}{\item[\textit{Acronyms}]}{%
 `\ifthenelse{\equal{#1}{V}}{\item[\textit{Variables and parameters}]}{}}}}}
\usepackage{scalerel}
\usetikzlibrary{svg.path}
\definecolor{orcidlogocol}{HTML}{A6CE39}
\tikzset{
 orcidlogo/.pic={
 \fill[orcidlogocol] svg{M256,128c0,70.7-57.3,128-128,128C57.3,256,0,198.7,0,128C0,57.3,57.3,0,128,0C198.7,0,256,57.3,256,128z};
 \fill[white] svg{M86.3,186.2H70.9V79.1h15.4v48.4V186.2z}
 svg{M108.9,79.1h41.6c39.6,0,57,28.3,57,53.6c0,27.5-21.5,53.6-56.8,53.6h-41.8V79.1z M124.3,172.4h24.5c34.9,0,42.9-26.5,42.9-39.7c0-21.5-13.7-39.7-43.7-39.7h-23.7V172.4z}
 svg{M88.7,56.8c0,5.5-4.5,10.1-10.1,10.1c-5.6,0-10.1-4.6-10.1-10.1c0-5.6,4.5-10.1,10.1-10.1C84.2,46.7,88.7,51.3,88.7,56.8z};
 }
}

\newcommand\orcidicon[1]{\href{https://orcid.org/#1}{\mbox{\scalerel*{ \begin{tikzpicture}[yscale=-1,transform shape]
 \pic{orcidlogo};
 \end{tikzpicture}
 }{|}}}}

\begin{document}
%
% paper title
\title{Can locational disparity of prosumer energy optimization due to inverter rules be limited?}
% \title{Robust chance constrained day-ahead flexibility 
% planning and
% needs assessment
% for distribution networks}

%% To specify the authors when (number of affiliations > 2)
% \author{\IEEEauthorblockN{Md Umar Hashmi,
% Arpan Koirala,
% Hakan Ergun, 
% Dirk Van Hertem}
% \IEEEauthorblockA{\textit{Electa-ESAT, KU Leuven \& EnergyVille},
% Genk, Belgium}
% \IEEEauthorblockA{(mdumar.hashmi, arpan.koirala, hakan.ergun, dirk.vanhertem)@kuleuven.be}
% % \IEEEauthorblockA{\IEEEauthorrefmark{3} ITI/LARSYS and prsma.com,
% % Funchal, Portugal}
% % \IEEEauthorblockA{\IEEEauthorrefmark{4}California Institute of Technology,
% % Passdena, CA, USA}
% % \IEEEauthorblockA{\IEEEauthorrefmark{5}EDF France, 
% % Paris, France}
% }
\author{Md~Umar~Hashmi$^{1}$,~\IEEEmembership{Member,~IEEE}~\orcidicon{0000-0002-0193-6703},~Deepjyoti~Deka$^{2}$,~\IEEEmembership{Senior~Member~IEEE}~\orcidicon{0000-0003-3928-3936},~Ana~Bu\v{s}i\'c$^{3}$,~\IEEEmembership{Member,~IEEE}~\orcidicon{0000-0002-4133-3739}, and~Dirk~Van~Hertem$^{1}$,~\IEEEmembership{Senior~Member,~IEEE}~\orcidicon{0000-0001-5461-8891}
\thanks{Corresponding author email: mdumar.hashmi@kuleuven.be}
\thanks{$^{1}$M.U.H. and D.V.H. are with KU Leuven, division Electa \& EnergyVille, Genk, Belgium}
\thanks{$^{2}$D.D. is with Los Alamos National Laboratory, USA }
\thanks{$^{3}$A.B is with INRIA, DI ENS, Ecole Normale Sup\'erieure, CNRS, PSL Research University, Paris, France.}%
% \thanks{R. D'hulst is with VITO \& EnergyVille, Genk, Belgium}
\thanks{This work is supported by the H2020 EUniversal project, grant agreement ID: 864334 (\url{https://euniversal.eu/}) and 
% Moonshot Catalisti projects (\url{https://catalisti.be/moonshot/})
% We gratefully acknowledge the financial support of 
the Flemish Government and Flanders Innovation \& Entrepreneurship (VLAIO) through the Moonshot project InduFlexControl (HBC.2019.0113).
}}

% make the title area
\maketitle

% As a general rule, do not put math, special symbols or citations
% in the abstract

% \tableofcontents

\begin{abstract}
To mitigate issues related to the growth of variable smart loads and distributed generation, distribution system operators (DSO) now make it binding for prosumers with inverters to operate under pre-set rules. In particular, the maximum active and reactive power set points for prosumers are based on local voltage measurements to ensure that inverter output does not cause voltage violations. However, such actions, as observed in this work, restrict the range available for local energy management, with more adverse losses on arbitrage profits for prosumers located farther away from the substation. 
{The goal of the paper is three-fold:
(a) to develop an optimal local energy optimization algorithm for activation of load flexibility and inverter-interfaced solar PV and energy storage under time-varying electricity prices; (b) to quantify the locational impact on prosumer arbitrage gains due to inverter injection rules prevalent in different energy markets; (c) to propose a {computationally efficient} hybrid inverter control policy which provides voltage regulation while substantially reducing locational disparity. Using numerical simulations on three identical prosumers located at different parts of a radial feeder, we show that our control policy is able to minimize locational disparity in arbitrage gains between customers at the beginning and end of the feeder to 1.4\%, while PV curtailment is reduced by 91.7\% compared to the base case {with restrictive volt-Var and volt-watt policy}.}
\end{abstract}

\tableofcontents


\begin{table}[!htbp]
% \centering
\small
\begin{tabular}{ll}
\normalsize{\textcolor{black}{\textit{Abbreviation}}} & \\
ANRC & Avoiding Negative Reinforcement Control \\
CVC & Cumulative Voltage Correction \\
DN & Distribution Network \\
DG & Distributed Generation \\
DSO & Distribution System Operator \\
% HEMS & Home Energy Management System \\
% FAS & Flexibility activation signal \\
HVAC & Heating, Ventilation, and Air Conditioning\\
% LMP & Locational marginal price \\
LCG & Loss of Consumer Gain\\
LP & Linear Programming \\
% LV & Low Voltage \\
PCC & Point of Common Coupling \\
PRC & Positive reinforcement control \\
P(U) & Volt-Watt \\
Q(U) & Volt-Var \\
PV & PhotoVoltaic\\
TCI & Total Curtailed Energy\\
VCI & Voltage Correction Index\\
VR & Voltage regulation
\end{tabular}
\end{table}

\pagebreak

\section{Introduction}
\textcolor{black}{The penetration of distributed generation (DG) and flexible loads in low voltage (LV) distribution networks (DN) is growing at an astounding pace, and near future projections point towards a substantial share of total energy consumed being met. Along with DG, prosumers with flexible loads such as water heaters, HVAC \cite{hassan2020hierarchical}, pool pumps \cite{meyn2015ancillary}, and energy storage can perform local energy optimization to minimize their electricity bill. Further, \textcolor{black}{distribution system operators (DSOs)} motivate such LV prosumers to be responsive by introducing time-varying electricity prices, net-energy metering, peak demand charge, etc. The authors in \cite{mathieu2013energy, hashmi2017optimal, hassan2018optimal} use thermostatically controlled load and energy storage for performing energy arbitrage, while \cite{engels2019optimal} uses energy storage for peak shaving and frequency control, and \cite{hashmi2020arbitrage} studies power factor correction along with arbitrage. The authors in \cite{hashemi2016efficient, bletterie2016voltage, lee2020maximizing} use energy storage for increasing \textcolor{black}{photovoltaic (PV)} hosting capacity. The growth of {DGs in a DN} causes several problems for DSO such as localized voltage rise beyond permissible limits, \textcolor{black}{and} reverse power flow causing damage to electrical appliances \cite{chmielowiec2021photovoltaic}. In order to mitigate these issues, DSOs choose either or a combination of four paths: (a) develop new inverter connection grid rules, (b) upgrade DN with reinforcements, such as installing tap changing transformers, etc, (c) curtail renewable generation or load in case of voltage rise/dip beyond thresholds, or (d) create a market for procuring prosumer load flexibility directly or through an aggregator. Indeed, IEEE-1547-2018 standard makes it mandatory for incoming DGs to comply with voltage regulation capabilities \cite{narang2019highlights, ieee2018ieee} to minimize system violations, as summarized in Table \ref{tab:ieee1547}. The standard promotes operation modes such as (a) constant power factor mode, (b) constant reactive power mode, (c) volt-Var, (d) P-Q mode, and (f) volt-watt mode.}

\begin{table}[!htbp]
\centering
\caption{\small{Desired capabilities according to IEEE-1547-2018 \cite{narang2019highlights, ieee2018ieee} }}
\label{tab:ieee1547}
\begin{tabular}{p{25mm}|p{15mm}|p{15mm}|p{19mm}|p{15mm}|p{17mm}} 
% \begin{tabular}{lllllll}
\hline
 & \multicolumn{5}{l|}{Mandatory voltage regulation capability modes} \\ 
\cline{2-6}
 {Performance Categories} & Constant power factor ($\cos \phi$) & Constant reactive power & Voltage-reactive power (volt-Var) & {Active power-reactive power} & {Voltage-active power (volt-watt)} \\ 
\hline
Category A: Minimum performance & {\cellcolor{cyan}}yes & {\cellcolor{cyan}}yes & {\cellcolor{cyan}}yes & Not required & Not required \\ 
\hline
Category B: for smart inverters & {\cellcolor{cyan}}yes & {\cellcolor{cyan}}yes & {\cellcolor{cyan}}yes & {\cellcolor{cyan}}yes & {\cellcolor{cyan}}yes \\
\hline 
\end{tabular}
\end{table}

In Europe, inverter control rules are also commonly used by the DSOs \cite{troester2009new,juamperez2014voltage}. This involves controlling the active (P) and reactive (Q) power \textcolor{black}{output} of the inverter as a function of local voltage measurements (U). This is also referred to as volt-watt and volt-Var control in literature \cite{braslavsky2017voltage}. As an example, \textcolor{black}{reactive power injection rules are shown in Table \ref {tab:mitnetz} for Mitnetz Strom, a DSO in Eastern Germany \cite{linkmitnetz}}
\begin{table}[!htbp]
\centering
\caption{\small{Mitnetz Strom inverter operational modes \cite{linkmitnetz}}}
\label{tab:mitnetz}
\begin{tabular}{l|l|l}
 & Inverter{ $\leq$ 4.6 kVA} & Inverter{ $\geq$ 4.6 kVA} \\
 \hline
Desired $\cos \phi$ & 0.95 & 0.9 \\ \hline
\multirow{3}{*}{Generation} & 1) $\cos \phi$ (P) characteristic & 1) Volt-Var \\
 & \multirow{2}{*}{2) Fixed $\cos \phi$} & 2) $\cos \phi$ (P) characteristic \\
 & & 3) Fixed $\cos \phi$ \\ \hline
\multirow{2}{*}{Storage} & \multirow{2}{*}{1) Fixed $\cos \phi$} & 1) Volt-Var \\
 & & 2) Fixed $\cos \phi$
 \end{tabular}
\end{table}

P(U) and Q(U) inverter control in standalone and/or in combination have been studied in \cite{fu2015optimal,weckx2014combined, karagiannopoulos2017hybrid, mak2020optimization} for performing localized voltage regulation, ensuring DN voltage does not aggravate due to additional injection or consumption of P and Q. The authors in \cite{olivier2015active, ustun2019impact, gebbran2021fair} use Q(U) control with active power curtailment as a last resort for mitigating overvoltages caused by PV injection. In \cite{cagnano2011online}, the authors propose reactive power control envelopes based on the unused solar inverter capacity for ensuring nodal voltages are within bounds. However, \cite{olivier2015active, cagnano2011online} do not consider local optimization of prosumers. Authors in \cite{dall2014optimal, ji2018centralized} propose centralized dispatch of PV inverters for avoiding voltage issues and minimizing curtailment. However, centralized control for small-sized inverters is not practical as it requires feedback from local measurements, which is ineffective in DNs due to the incomplete spread of smart meters \cite{bhela2017enhancing} and privacy concerns \cite{knyrim2011smart}. This motivates us to focus on studying distributed inverter control with optimizing flexible prosumer's energy cost.

\textcolor{black}{In recent years, many works have highlighted the disparity caused in inverter usage due to prosumer location in a radial DN. \cite{sousa2020pv} proposes a PV hosting capacity mechanism in presence of inverter control policies such as volt-var and volt-watt, where they exemplify the locational impact of prosumer PV connection. Authors in \cite{zhan2015relay} underscore that the DN relay settings are affected by the location of DN protection relay in a radial DN. They utilize relay settings as a basis for DG placement. In \cite{choi2020study}, an assessment is performed to quantify the locational impact on the lifetime of PV inverters. The authors conclude that the operational life of the inverter is significantly affected by the installation site along with ambient temperatures. In radial DN, the voltage levels {drop} as one moves farther away from the feeder head or substation \cite{tonkoski2012impact, kabir2014improving}, leading to greater voltage fluctuations at farther locations. Hence, current local inverter rules will lead to different operational regimes for prosumers at different locations and affect their arbitrage opportunity. In this work, we will analyze locational impacts on load flexibility and arbitrage due to volt-Var and volt-watt modes of inverter control in detail. To this end, we design new hybrid modes for inverter control that remedy the locational disparity. It is worth noting that our study is in line with the mandate that DSOs should broadly provide a level-playing field for all prosumers consuming or injecting electrical energy, irrespective of prosumer location \cite{EU2019}.}

% \vspace{-5pt}

\begin{figure}[!htbp]
	\center
	\includegraphics[width=0.8\linewidth]{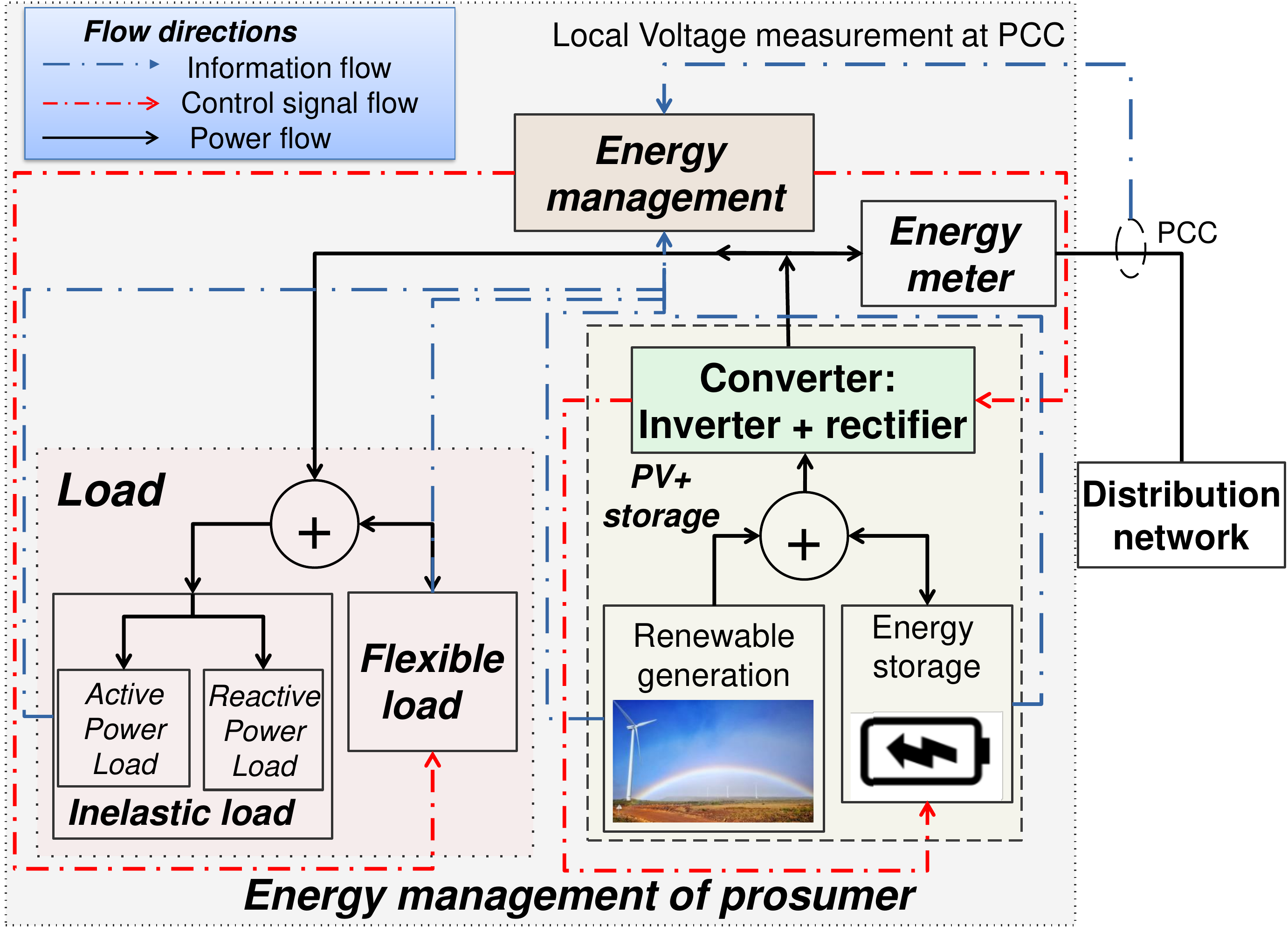}
% 	\vspace{-12pt}
	\caption{\textcolor{black}{Prosumer energy management with voltage regulation at the point of common coupling (PCC) using local voltage measurement}}
	\label{fig:systemnet}
\end{figure}

\textcolor{black}{\textit{Contributions}: 
The goal of the paper is to quantify the locational discrepancy for a prosumer with DG, energy storage, and flexible load due to contemporary DSO rules which enforce active and reactive power limits on inverter-interfaced generation. The prosumer considered in this paper is shown in Fig.~\ref{fig:systemnet}. } 
% \vspace{-5pt}
\textcolor{black}{The main contributions of the paper are:
\begin{itemize}
\item To evaluate the economic value of energy storage and load flexibility, using an LP (linear programming) based formulation for resource dispatch and further to quantify the locational disparity of prosumers that perform local energy optimization while following inverter control rules that ensure grid voltages are within the safe operational region.
\item To propose a novel local hybrid inverter control policy that minimizes the locational disparity {and is computationally efficient}. We compare our approach against two traditional inverter policies: (a) positive reinforcement control (PRC) and (b) avoiding negative reinforcement control (ANRC), which represent an optimistic and pessimistic interpretation of volt-Var and volt-watt control policies.
\end{itemize}}

Using the proposed framework, we observe the loss of consumer profit\footnote{Profit of a prosumer refers to the avoided cost due to energy optimization.} due to inverter control and note that a prosumer at the end of the feeder may have to pay more than 43\% in the variable component of their electricity bill, compared to a similar prosumer located near the distribution substation. Moreover, we observe that passive inverter rules may fail if a large capacity energy storage device is connected at a node, as storage can reverse the mode of operation, i.e. from charging to discharging and vice versa, which may cause voltage violation in the opposite direction compared to the direction of correction. Crucially, we observe that \textit{a hybrid inverter control policy where active power (P) is controlled using ANRC and reactive power (Q) is controlled using PRC significantly reduces the prosumer excess cost of consumption while ensuring the correction of nodal voltage.} 

This paper is structured as follows. In Section~\ref{section2}, the mathematical formulation for scheduling load flexibility and energy storage is performed using linear programming. This formulation is an extension of prior work on storage performing arbitrage proposed in \cite{hashmi2019optimal}. Section~\ref{section3} translates the inverter control rules into permissible ranges for an active and reactive generation. These ranges are utilized to validate the energy optimization output at a faster timescale. Section~\ref{section4} {details} the effect of the location of prosumer on DN and performance indices used, and directions in which fairness can be \textcolor{black}{incorporated}. Section~\ref{section5} presents the numerical results. Section~\ref{section6} concludes this paper.

\pagebreak

\section{Prosumer energy management}
\label{section2}
We consider a prosumer with inelastic and flexible load components, local generation, and energy storage (as shown in Fig.~\ref{fig:systemnet}). It is connected to the DN, from where it can buy or to which it can sell energy.
Based on buying and selling price fluctuations, load, and generation variations, the prosumer energy management system optimizes battery states and flexible loads at regular intervals to minimize the cost of energy. 
Moreover, the prosumer is obliged to follow active and reactive power injection rules as detailed in Section~\ref{section3}. 
% and enforced by the DSO. 
We now describe the optimization problem for the prosumer in detail. 

\subsection{Notation and system model}
 The price of electricity at time instant $i$ consists of $p_b^i$ (the buying price) and $p_s^i$ (the selling price). The difference between buying and selling prices is common in DSOs \cite{gautier2019self, hashmi2020storage}, and their ratio is denoted as $\kappa_i $. The end user's inelastic consumption is denoted as $d_i \geq 0$, the flexible load is denoted as $y_i \geq 0$, and renewable generation is given as $r_i \geq 0$. 
Net uncontrolled power seen at the energy meter is denoted as $z_i = d_i - r_i ~ \in \mathbb{R} $. 

\subsubsection{Timescale and notation}
Prosumer energy optimization is performed at a slower timescale at every time instant $i$.
The total time duration, $T$, 
of operation is divided into $N$ equal steps, indexed by $\{1,...,N\}$. 
The time duration of each step $1\leq i \leq N$ is denoted as $h$. Hence, $T=N h$.

\begin{figure}[!htbp]
	\center
	\includegraphics[width=4.7in]{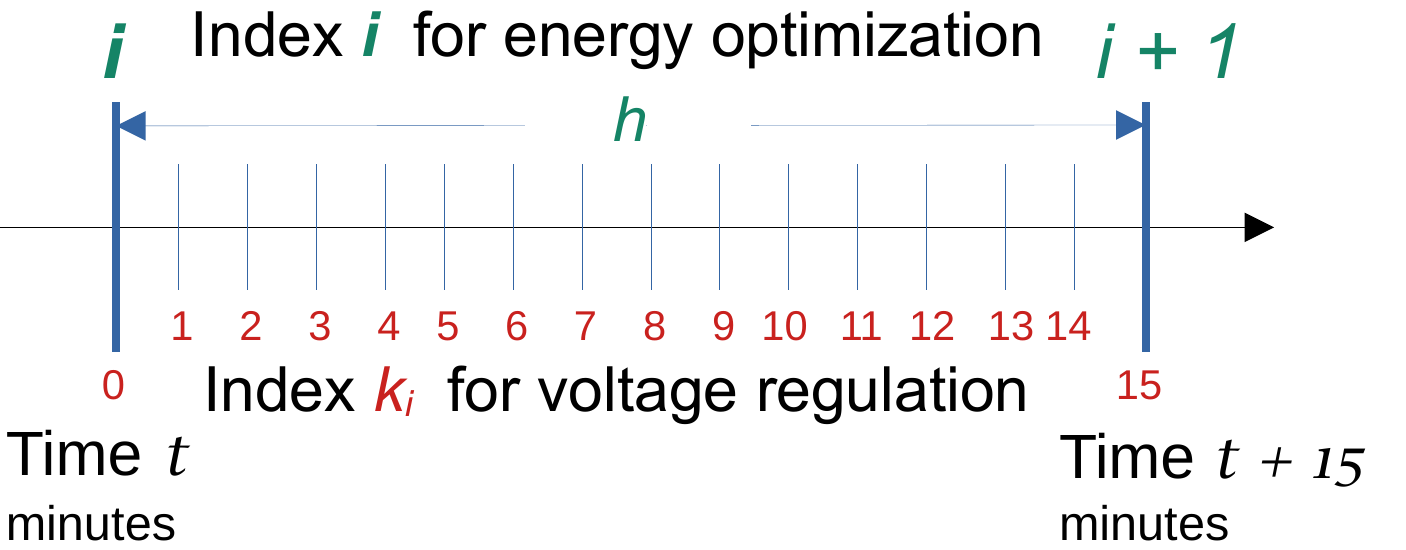}
	\vspace{-2pt}
	\caption{Pictorial representation of timescale for optimization and voltage regulation}
	\label{fig:timescale}
\end{figure}

The time period between $i$ and $i+1$ can be divided into a faster timescale as shown in Fig.~\ref{fig:timescale} and referred to using $k_i$. The value of $k_i$ resets to 0 at $i$.
At this faster timescale, local voltage regulation is performed.

In this work, energy management is performed every 15 minutes, and the voltage is regulated every minute based on the voltage measurement at PCC. The grid voltage at PCC is measured every minute. Active and reactive power outputs are adjusted to satisfy the grid's needs.

\subsubsection{Flexibility model}
The flexible component of the load can be controlled within a range while ensuring the cumulative energy consumed is not decreased. This is \textcolor{black}{given} by
\begin{equation}
 K - \epsilon \leq h \sum_{i=1}^N y_i \leq K +\epsilon, \text{and}~ y_i \in [y_{\min}^i,y_{\max}^i], ~ \forall i,
\label{eq:loadflexibility}
\end{equation}
where $K$ denotes targeted cumulative energy consumed by flexible loads (ensuring the quality of service). The flexible loads can be operated with an upper and lower envelope denoted as $y_{\max}^i$ and, $y_{\min}^i$ respectively. Note that the flexibility is derived from loads, therefore, $y_{\min}^i \geq 0$. \eqref{eq:loadflexibility} will ensure arbitrage benefits are due to energy management and not because of a reduction in total energy consumption.
% \textcolor{black}
{$\epsilon$ denotes a small number for ensuring total energy consumed by flexible devices is approximately equal to $K$ \textcolor{black}{with} a small slack. In the absence of slack, \eqref{eq:loadflexibility} is an equality constraint.}

\subsubsection{{Battery model}}
The battery model considers the ramping constraint, and the capacity constraint along with charging and discharging efficiencies denoted 
by $\eta_{\text{ch}}, \eta_{\text{dis}} \in (0,1]$, respectively. 
The energy optimization considers the change in energy levels of the battery \textcolor{black}{at time $i$} is denoted as $x_i$. Selection of $x_i$ as the decision variable ensures that the energy storage arbitrage problem is convex, provided \textcolor{black}{the ratio of selling and buying price of electricity denoted as $\kappa$ satisfies} $\kappa \leq 1$ \cite{hashmi2017optimal, hashmi2019optimal, hashmi2020storage}. Change in battery energy level at $i$ is defined as $x_i = h \delta_i$, where $\delta_i \in [\delta_{\min}, \delta_{max}]$ $\forall i$ denotes ramp rate of the battery. $\delta_i> 0 $ when the battery is charging and vice versa. Note, $\delta_i$ is in units of power and $x_i$ is in units of energy.
The battery charge level is denoted as
\begin{equation}
b_i = b_{i-1} + x_i, \quad b_i\in [b_{\min},b_{\max}], \forall i,
\end{equation}
where $b_{\min}, b_{\max}$ are the minimum and maximum battery capacity. The power consumed by a battery at time $i$ is denoted as
\begin{equation}
f(x_i)= \frac{[x_i]^+}{h \eta_{\text{ch}}} - \frac{\eta_{\text{dis}}[x_i]^-}{h}=\frac{1}{h\eta_{\text{ch}}}\max(0,x_i) - \frac{\eta_{\text{dis}}\max(0,-x_i)}{h}, 
\label{finverse}
\end{equation}
where $x_i$ must lie in the range from $X_{\min}=\delta_{\min}h$ to $X_{\max}={\delta_{\max}h}$.
The total power consumed between time step $i$ and $i+1$ is given as $L_i = z_i+y_i +f(x_i)$, where $z_i$ is the uncontrolled net load, $y_i$ is the flexible load, and $f(x_i)$ is the battery consumption. 

The battery model is denoted as $x$C-$y$C: the battery will require $1/x$ hours
to fully charge and $1/y$ hours to fully discharge.

\subsection{{Price based energy arbitrage}}
The optimal arbitrage problem is defined as the minimization of the 
cost of consumption of energy (sum of net inflexible load, flexible load, and storage) over a time horizon considering battery constraints and load flexibility constraints.
\begin{subequations}
\begin{align}\label{arbitrage_opt}
 &\min_{x,y} h\sum_{i}^N [ z_i + y_i +f(x_i) ]^+ p_{\text{b}}^i- [ z_i + y_i +f(x_i) ]^- p_{\text{s}}^i,\\
 &\text{s.t.} ~~\eqref{eq:loadflexibility}, \\
 & b_{\min} - b_0 \leq h \sum_i^N f(x_i) \leq b_{\max} - b_0, ~ \forall i, \\
 & x_i \in [X_{\min},X_{\max}], ~ \forall i. 
 \end{align}
\end{subequations}
The first constraint relates to load flexibility, the second to battery capacity, and the third to battery ramping. This formulation can be solved as a linear programming (LP) formulation, denoted as $(P_{LP})$ in Appendix~\ref{appendix:lpmatrix}. In the next section, we discuss the introduction of non-linear inverter size constraints that will be updated at a finer timescale based on voltage measurement. 

\pagebreak

\section{Inverter control and operation}
\label{section3}
\subsection{Inverter model}
The inverter, shared by PV and battery, has a maximum $S_{\max}$ rating in Volt-Ampere (VA). Its output over each 15-minute slot, based on energy optimization in \eqref{arbitrage_opt}, is denoted as $P_{\text{inv}}^i= f(x_i) - r_i$, \textcolor{black}{where $r_i$ denotes the renewable generation}. Within each time-slot $i$, the inverter active power $P_{\text{inv}}^{k_i}$ and the storage output $P_B^{k_i}$ are modeled at a faster timescale of 1 minute (see Fig.~\ref{fig:timescale}), related as 
\begin{align}\label{inverter_eq}
 P_{\text{inv}}^{k_i}= P_B^{k_i} - r_i+ P_{\text{curt}}^{k_i},
\end{align}
where $P_{\text{curt}}^{k_i}$ denotes curtailed active power. Note $P_{\text{curt}}^{k_i} \in [0,r_i]$. The inverter power limits are given as
\begin{subequations}
\label{eq:upperinverterbounds}
\begin{equation}
 P_{\text{inv}}^{k_i} \in [-P_{\max}, P_{\max}], \text{ where } P_{\max} \leq S_{\max}, 
\end{equation}\vspace{-5pt}
\begin{equation} \begin{split}
&Q_{\text{inv}}^{k_i} \in [-Q_{\max}, Q_{\max}] \text{ where } \\
&\begin{cases}
Q_{\max}=P_{\text{inv}}^{k_i} \tan(\cos^{-1}{\text{pf}^{wc})}, & \text{if } P_{\max} \in [0.1,\text{pf}^{wc}S_{\max}], \\ 
Q_{\max}= \sqrt{(S_{\max})^2 - (P_{\text{inv}}^{k_i})^2 }, & \text{otherwise,} 
\end{cases}
\end{split}
\label{eq:upperinverterboundsQ}
\end{equation} 
\end{subequations}
Here, $\text{pf}^{wc}$ denotes the worst-case power factor set by the DSO. \begin{figure}[!htbp]
	\center
	\includegraphics[width=3in]{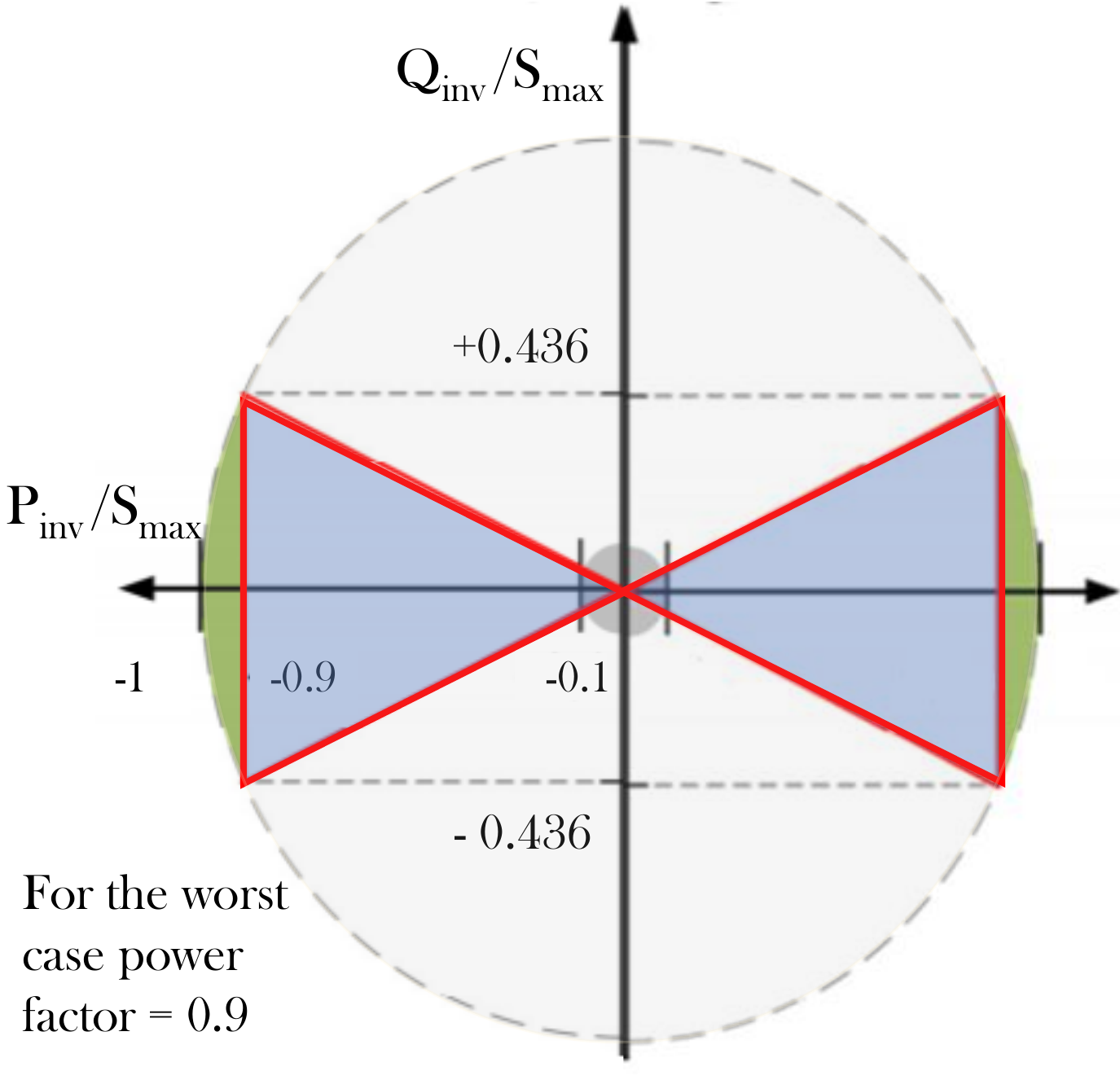}
% 	\vspace{-10pt}
	\caption{P, Q feasible region of inverter based on power factor limit of 0.9}
	\label{fig:inverterfes}
\end{figure}
Fig.~\ref{fig:inverterfes} denotes the feasible active and reactive power regions based on the worst-case power factor limit. The blue-shaded and green regions are where the inverter is allowed to operate. 

We next discuss the inverter rules for P(U) and Q(U) control, implemented at a faster timescale of one minute. The DSO imposes such control based on local voltage $U$. Many recent works such as \cite{jabr2019robust, padullaparti2016advances, braslavsky2015stability, chathurangi2021comparative, olowu2021optimal} have explored similar volt-var (Q(U)) and volt-watt (P(U)) inverter control policy design. As standard, we consider inverter control to operate with \textit{P-priority}, i.e., priority being given to active power output in case both P, Q set-points \textcolor{black}{cannot} be met. 
\begin{figure*}[!htbp]
	\center
	\includegraphics[width=6.64in]{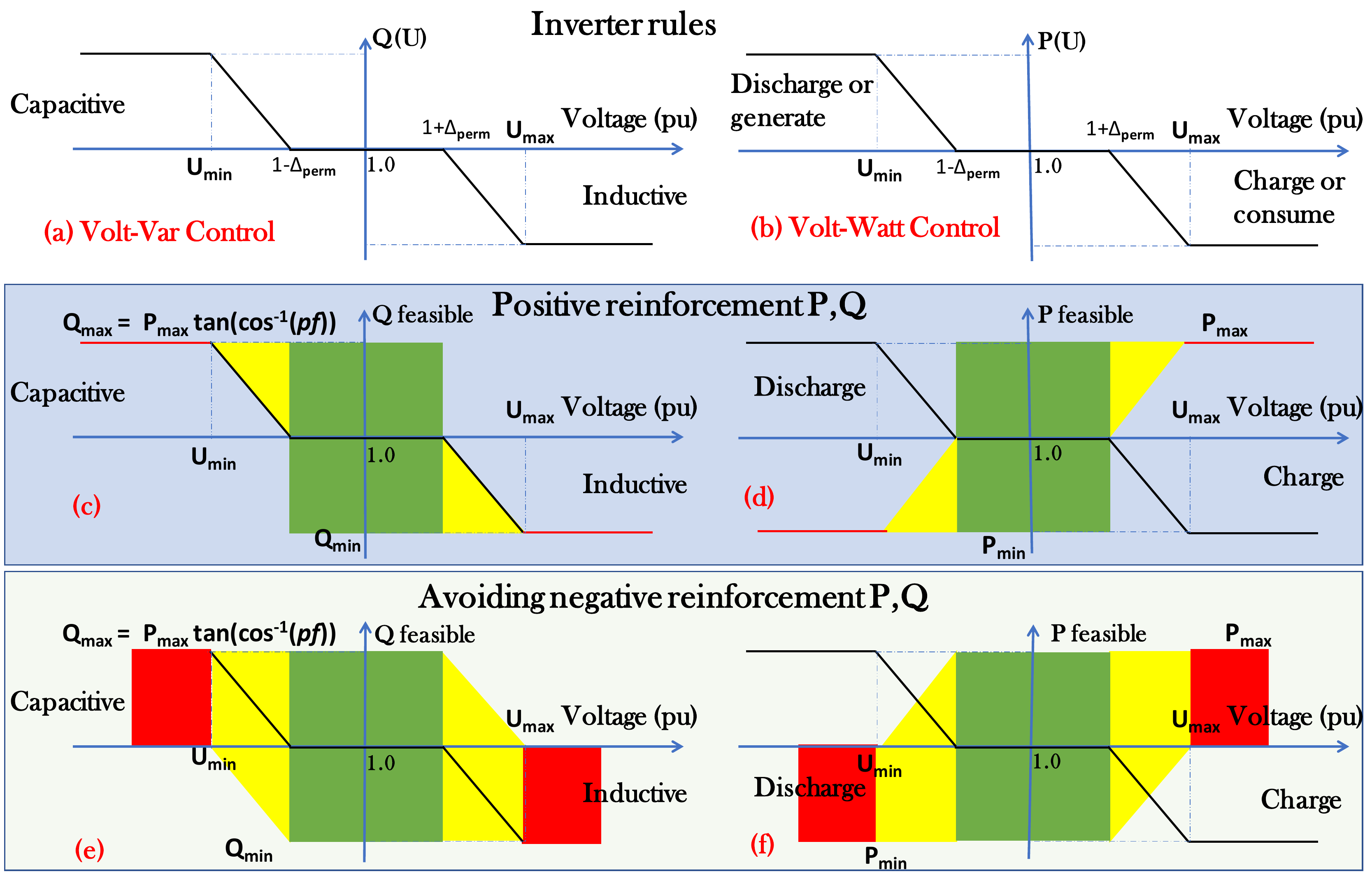}	
\caption{(a) Volt-var or Q(U) control and (b) volt-watt or P(U) translated into energy storage inverter feasible operation region considering \textit{positive reinforcement} towards voltage regulation at the node denoted in the green region in (c) and (d). (e) and (f) denotes feasible region for \textit{avoiding} \textit{negative reinforcement} towards voltage regulation at the node. The feasible region is depicted as traffic-light control with green as no reduction in the feasible region, the yellow region shrinks the feasible region, and the red region shrinks it even further.}\label{fig:feasiblepqv}
\end{figure*}
% \vspace{-5pt}

\subsection{Voltage zones for inverter control}
\label{subsec31}
Consider the instantaneous local voltage magnitude $U_{k_i}$ at PCC at time $k_i$. As shown in Fig.~\ref{fig:feasiblepqv} (a) and (b), the operational voltage can be divided into 5 zones:
\begin{itemize}
 \item Zone 1: $U_{k_i} <U_{\min}$,
 \item Zone 2: $U_{k_i} \in [U_{\min}, 1-\Delta_{perm})$,
 \item Zone 3: $U_{k_i} \in [1-\Delta_{perm}, 1+\Delta_{perm}]$,
 \item Zone 4: $U_{k_i} \in (1+\Delta_{perm}, U_{\max}]$,
 \item Zone 5: $U_{k_i} > U_{\max}$,
\end{itemize}
where $\Delta_{perm}$ denotes the permissible voltage deviation in the distribution network before any voltage regulation is required.
The values of $U_{\min}$, $U_{\max}$ are defined by the DSO. The value of $\Delta_{\text{perm}}$ will decide the droop slopes. For simplicity, we assume $\Delta_{\text{perm}}$ to be symmetrical around under and over-voltage. 

The operating region for the inverter is defined based on the inverter's characteristics and instantaneous voltage $U$. For active power, we denote the region as $R_P^U$, and for reactive power, as $R_Q^U$.
\begin{equation}\label{R^P_defn}
\begin{split}
 R_P^U(U_{k_i}) \equiv [ R^U_{P_{\min}}, R^U_{P_{\max}}], \\ R_Q^U(U_{k_i}) \equiv [R^U_{Q_{\min}}, R^U_{Q_{\max}}],
\end{split}
\end{equation}
where $ R^U_{P_{\min}}$, $ R^U_{Q_{\min}}$ denote the lower operating envelopes and $ R^U_{P_{\max}}$ and $ R^U_{Q_{\max}}$ denote the upper operating envelopes. Note that these ranges are modified based on the zone that the voltage exists in and the kind of control. In the next sections \ref{subsec32} and \ref{subsec33}, we define three such control designs. 
\subsection{PRC inverter operation}
\label{subsec32}
Under the \textit{Positive reinforcement control} (PRC) model for P, Q, the inverter actively contributes to rectifying voltage issues in the different nodal voltage zones described in Section \ref{subsec31}. The ranges for P and Q under PRC are listed in Table~\ref{tab:positiveLimits} and depicted in Figs.~\ref{fig:feasiblepqv} (c) and (d). For extreme nodal voltage levels (zones Z1 or Z5), the operating region is just one point. For moderate voltages (Z2 and Z4), the ranges for inverter output are restricted for voltage regulation using a linear droop. Finally, for voltage in Z3, inverter control does not limit arbitrage optimization opportunities.
\vspace{-12pt}
\begin{center}
\begin{table}[!ht]
\centering
\small
	\caption {Positive reinforcement inverter mode of operation} %\vspace{-12pt}
	\label{tab:positiveLimits}
	\begin{center}
		\begin{tabular}{|p{0.3cm}|p{4.75cm}|p{4.8cm}|}
			\hline
			{\textbf{$U_{k_i}$}} & $[R^U_{P_{\min}}, R^U_{P_{\max}}]$ &$[R^U_{Q_{\min}}, R^U_{Q_{\max}}]$\\[1mm]
			\hline
			\colorbox{red}{Z1} & [$P_{\min}$, $P_{\min}$] & [$Q_{\max}$, $Q_{\max}$]\\
 \colorbox{yellow}{Z2} & [$P_{\min}$, $\frac{P_{\min}(U_{k_i} - (1- \Delta_{\text{perm}}))}{U_{\min} - (1- \Delta_{\text{perm}})} $] & [$\frac{Q_{\max}(U_{k_i} - (1- \Delta_{\text{perm}}))}{U_{\min} - (1- \Delta_{\text{perm}})} $, $Q_{\max}$]\\ [1mm]
 \colorbox{green}{Z3} & [$P_{\min}$, $P_{\max}$] & [$Q_{\min}$,$Q_{\max}$]\\ 
 \colorbox{yellow}{Z4} & [$\frac{P_{\max}(U_{k_i} - (1+ \Delta_{\text{perm}}))}{U_{\max} - (1+ \Delta_{\text{perm}})} $, $P_{\max}$] & [$Q_{\min}$, $\frac{Q_{\min}(U_{k_i} - (1+ \Delta_{\text{perm}}))}{U_{\max} - (1+ \Delta_{\text{perm}})} $]\\ 
 \colorbox{red}{Z5} & [$P_{\max}$, $P_{\max}$] & [$Q_{\min}$, $Q_{\min}$]\\ 
 \hline
		\end{tabular}
		\hfill\
	\end{center}
\end{table}
% \addtocounter{footnote}{-2}
\end{center}
% \vspace{-12pt}

\subsection{ANRC inverter operation}
\label{subsec33}
Unlike PRC, \textit{avoiding negative reinforcement control} (ANRC) model does not make the inverter actively contribute to mitigating voltage problems but prevents it from aggravating the current condition. The ranges for P and Q under ANRC are provided in Table ~\ref{tab:negativeLimits} and Figs.~\ref{fig:feasiblepqv} (e) and (f). Note that the $R^U$ for P, Q for ANRC is larger than or equal to the range generated using PRC, therefore, ANRC is less restrictive for prosumer energy optimization compared to PRC.
\begin{center}
\begin{table}[!h]
\centering
\small
	\caption {ANRC inverter mode of operation} %\vspace{-8pt}
	\label{tab:negativeLimits}
	\begin{center}
		\begin{tabular}{|p{0.3cm}|p{4.5cm}|p{4.3cm}|}
			\hline
			{\textbf{$U_{k_i}$}} & $[R^U_{P_{\min}}, R^U_{P_{\max}}]$ &$[R^U_{Q_{\min}}, R^U_{Q_{\max}}]$ \\[1mm]
			\hline 
			\colorbox{red}{Z1} & [$P_{\min}$, 0] & [0, $Q_{\max}$]\\
 \colorbox{yellow}{Z2} & [$P_{\min}$, $\frac{P_{\max}(U_{\min} - U_{k_i})}{U_{\min} - (1- \Delta_{\text{perm}})} $] & [$\frac{Q_{\min}(U_{\min} - U_{k_i})}{U_{\min} - (1- \Delta_{\text{perm}})} $, $Q_{\max}$]\\ [1mm]
 \colorbox{green}{Z3} & [$P_{\min}$, $P_{\max}$] & [$Q_{\min}$, $Q_{\max}$]\\ 
 \colorbox{yellow}{Z4}& [$\frac{P_{\min}(U_{\max} - U_{k_i})}{U_{\max} - (1+ \Delta_{\text{perm}})} $, $P_{\max}$] & [$Q_{\min}$, $\frac{Q_{\max}(U_{\max} - U_{k_i})}{U_{\max} - (1+ \Delta_{\text{perm}})} $]\\ 
 \colorbox{red}{Z5} & [0, $P_{\max}$] & [$Q_{\min}$, 0]\\ 
 \hline
		\end{tabular}
		\hfill\
	\end{center}
\end{table}
% \addtocounter{footnote}{-2}
\end{center}
% \vspace{-10pt}
\subsection{Hybrid inverter operation}
From Fig.~\ref{fig:feasiblepqv}, we observe that PRC is more restrictive for active and reactive power control compared to ANRC. Typically, energy management in low voltage consumers deals with only active power, therefore, PRC-based inverter P control can lead to a significant reduction in benefits for prosumer energy optimization. We propose a hybrid inverter control policy that selects active power output based on ANRC and reactive power output based on PRC. The lower and upper bound of ranges for this based on voltage magnitude is listed in Table~\ref{tab:hybridinverter}.
\begin{center}
\begin{table}[!h]
\centering
\small
	\caption {Hybrid inverter mode of operation} %\vspace{-8pt}
	\label{tab:hybridinverter}
	\begin{center}
		\begin{tabular}{|p{0.3cm}|p{4.5cm}|p{4.85cm}|}
			\hline
			{\textbf{$U_{k_i}$}} &$[R^U_{P_{\min}}, R^U_{P_{\max}}]$ &$[R^U_{Q_{\min}}, R^U_{Q_{\max}}]$ \\[1mm]
			\hline 
			\colorbox{red}{Z1} & [$P_{\min}$, 0] & [$Q_{\max}$, $Q_{\max}$]\\
 \colorbox{yellow}{Z2} & [$P_{\min}$, $\frac{P_{\max}(U_{\min} - U_{k_i})}{U_{\min} - (1- \Delta_{\text{perm}})} $] & [$\frac{Q_{\max}(U_{k_i} - (1- \Delta_{\text{perm}}))}{U_{\min} - (1- \Delta_{\text{perm}})} $, $Q_{\max}$]\\ [1mm]
 \colorbox{green}{Z3} & [$P_{\min}$, $P_{\max}$] & [$Q_{\min}$, $Q_{\max}$]\\ 
 \colorbox{yellow}{Z4}& [$\frac{P_{\min}(U_{\max} - U_{k_i})}{U_{\max} - (1+ \Delta_{\text{perm}})} $, $P_{\max}$] & [$Q_{\min}$, $\frac{Q_{\min}(U_{k_i} - (1+ \Delta_{\text{perm}}))}{U_{\max} - (1+ \Delta_{\text{perm}})} $]\\ 
 \colorbox{red}{Z5} & [0, $P_{\max}$] & [$Q_{\min}$, $Q_{\min}$]\\ 
 \hline
		\end{tabular}
		\hfill\
	\end{center}
\end{table}
% \addtocounter{footnote}{-2}
\end{center}

\subsection{Minimizing Constraint validation}
\label{subsec34}
From \eqref{inverter_eq}, solving $P_{LP}$ (energy arbitrage) without any inverter control gives $P_{inv}^{k_i} = P^i_{inv} = f(x_i) - r_i$ with $P_{\text{curt}}^{k_i} = 0$, and 
 $Q_{inv}^{k_i} = Q_{\text{default}} = 0 ~\forall ~k_i$. Under the permissible active P(U) and reactive limits Q(U) defined under PRC, ANRC, or hybrid control, the following \textcolor{black}{three} conditions may emerge:
\begin{itemize}
 \item $P^i_{inv}$ satisfies both the output constraints derived by P(U) and Q(U) curves,
 \item $P^i_{inv}$ satisfies only one of the output constraints derived by P(U) or Q(U),
 \item $P^i_{inv}$ does not satisfy both output constraints derived by P(U) and Q(U) curves.
\end{itemize}
For cases where the output $P^i_{inv}$ lies outside the feasible range $[R^U_{P_{\min}}, R^U_{P_{\max}}]$ of voltage-based operation, let the nearest boundary to $P^i_{inv}$ be $P^U_{trgt}$. The controller solves the following linear programming optimization problem ($P_{LP}^{curt}$) to determine minimum curtailment of $r_i$ to reach $P^U_{trgt}$: 
\begin{subequations}\label{eq:findingcurtailment}
\begin{align}
 (P_{LP}^{curt})~~\min_{P_{B}^{k_i}}~~ &P_{\text{curt}}^{k_i}\\
\text{s.t.}~&P_B^{k_i} -(r_i- P_{\text{curt}}^{k_i}) = P^U_{trgt}\\
 &0 \leq P_{\text{curt}}^{k_i} \leq r_i,\\
 &P_B^{k_i} \in \Big[\max\Big(\delta_{\min}, \frac{(b_{\min}-b_{k-1})\eta_{\text{dis}}}{k} \Big),\nonumber\\ &\hspace{30pt}\min\Big({\delta_{\max}}, \frac{(b_{\max}-b_{k-1})}{k\eta_{\text{ch}}} \Big) \Big].
\end{align}
\end{subequations}
Here, $P^U_{trgt}$ is based on the corresponding zone in Tables \ref{tab:positiveLimits}, \ref{tab:negativeLimits}, or \ref{tab:hybridinverter} and voltage $U_{k_i}$. Note that in some cases, ($P_{LP}^{curt}$) may not have a feasible solution due to too low or high state of charge of the battery, high renewable generation, or limited feasible range of operation for the inverter. In that case, $P_{\text{curt}}^{k_i}$ and $P_B^{k_i}$ are operated based on sign of $P^U_{trgt}$, i.e., if $P^U_{trgt}>0$ then charge the battery at
$P_B^{\max^{k_i}}$, else discharge at $P_B^{\min^{k_i}}$.
If $P^U_{trgt}>0$ then $P_{\text{curt}}^{k_i}$ is set at
$r_i$, else it is set to $0$. The inverter's remaining capacity is then utilized to set the reactive power. \textcolor{black}{In case the remaining capacity is not enough, the reactive power is set closest to the minimum or maximum bound level based on Tables~\ref{tab:positiveLimits}, \ref{tab:negativeLimits}, or \ref{tab:hybridinverter}. The inverter control procedure is detailed in Appendix \ref{appendix:control_algo}.} 

Algorithm \ref{alg:homeenergymanagement} details the entire procedure, conducted in a receding horizon fashion. Here the energy arbitrage is first solved using $(P_{LP})$ at Step 4 to give battery states at a slower timescale. Then the voltage is regulated based on inverter rules at a faster timescale for each time slot. First, the inverter rules for selected permissible P and Q ranges are identified in Step 10. Then the curtailed active power and reactive injection are determined using the associated Algorithm \ref{alg:control}, detailed in Appendix \ref{appendix:control_algo}. The battery's initial state for the next 15 minutes is fixed based on the battery charge level at the end of the faster timescale.

\begin{algorithm}
	\small{\textbf{Inputs}: $U_{k_i}, r_i, b_0, S_{\max}$, battery parameters}
	\begin{algorithmic}[1]
	 \State Set $P_{\max} = S_{\max}$, $i=1$
	 \vspace{2pt}
		\While{$i <= N$}
		\State Based on battery parameters, price levels, inelastic load, flexible load, renewable generation, initial battery level calculated matrices $A,b,X, f, lb, ub,$ detailed in Appendix \ref{appendix:lpmatrix},
		\State Solve $P_{LP}$ using load, generation, electricity price values, and initial battery capacity,
		\State Set $k=0$
		\While{$k < 15$}
% 		\State Implement local optimization every 15 min to find out $f(x_i)$,
		\State Increment $k=k+1$,
	 \State Inverter output based on energy management $\zeta_i=f(x_i) - r_i$,
	 % \State Initialize, 
		\State Measure local voltage $U_{k_i}$,
		\State Based on $U_{k_i}$ calculate $R^U_{P_{\min}}, R^U_{P_{\max}}, R^U_{Q_{\min}}, R^U_{Q_{\max}}$ as global variables. Refer to Table~\ref{tab:positiveLimits} for PRC, Table~\ref{tab:negativeLimits} for ANRC and
		Table~\ref{tab:hybridinverter} for hybrid inverter control,
		\State Calculate $Q_{\max}$ based on \eqref{eq:upperinverterboundsQ} as a function of $R_P^U$ %\textcolor{red}{dont call it $R_P^U$ that is the range of P. it think it should be $R^U_{P_min}$}
		%\textcolor{black}{this is P-priority}
	 \State Execute Algorithm \ref{alg:control} with range derived from Tables~\ref{tab:positiveLimits}, \ref{tab:negativeLimits}, or \ref{tab:hybridinverter} to find $P_{\text{curt}}^{k_i}, P_{\text{inv}}^{k_i}, Q_{\text{inv}}^{k_i}$,
	 \State Update battery charge level $b_{k_i}=b_{{k-1}_i} + f^{-1}(P_B^{k_i})$
	 \State Concatenate results $P_{\text{curt}}, P_{\text{inv}}, Q_{\text{inv}}, b_{{k=15}_{i}}$.
		\EndWhile 
	 \State Update battery charge level $b_i=b_{{k=15}_i}$ %\textcolor{red}{there is no 15 }
		\State Increment $i=i+1$,
		\State Calculate matrices $A,b,X, f, lb, ub$ for time $i$ till $N$ 
		\EndWhile 
% 		\State Calculate loss of consumer gain (LCG), total curtailed energy (TCE), Voltage correction index (VCI), cumulative voltage correction (CVC).
		\State Return $P_{\text{curt}}, P_{\text{inv}}, Q_{\text{inv}}$.
	\end{algorithmic}
	\caption{\texttt{Prosumer energy management with inverter rules}}\label{alg:homeenergymanagement}
\end{algorithm}
In the next section, we describe methods to understand the effect of location on the prosumer operation, while imposing the inverter controls with arbitrage.

\pagebreak

\section{Effect of prosumer location on DN feeder}
\label{section4}
In this section, we develop performance indices for understanding the effect of prosumer location {in} a DN on its optimization for arbitrage, with rules for DN voltage correction, and the effect of PRC, ANRC, or hybrid voltage control policies.
\vspace{-8pt}
\subsection{Performance indices}
\textcolor{black}{The proposed indices relate to (a) voltage violations (b) arbitrage gains (c) energy curtailment \cite{yamane2019determination, weckx2014combined}, all of which are very relevant to operational issues in the distribution grid as well as understanding the extent of profitability and opportunity cost associated with inverter interfaced resources in the distribution grids.  In fact, the comparison of arbitrage profits versus issues of grid limits on voltage is the primary objective of study for several smart grid control/optimization algorithms proposed in the literature as mentioned in \cite{navidi2019co}.
% \textcolor{red}{need to add literature}
}

We consider the following performance indices:\\
\textbf{Prosumer energy bill:} If inverter rules and associated control are imposed on prosumers, their savings due to the operation of the battery will be reduced due to voltage fluctuations, with more loss at nodes with greater fluctuation of voltage such as terminal nodes. Note that the cost of consumption without inverter control at node $j$ is denoted as 
\begin{equation}
 C_{\text{wic}}^j = \sum_i \Big(p_{b}^i (d_i+y_i -r_i +f(x_i))^+ - p_{s}^i (d_i+y_i -r_i +f(x_i))^-\Big)h.
 \label{eq:invperfindx}
\end{equation}
The cost of consumption with inverter control for a prosumer connected at node $j$ is denoted as
\begin{align}
 C_{\text{inv}}^j = \sum_i^N &p_{b}^i \big(d_i+y_i -r_i +\frac{1}{15}\sum_{k=1}^{15} (P_B^{k_i}+P_{\text{curt}}^{k_i})\big)^+h\nonumber\\
 -&p_{s}^i \big(d_i+y_i -r_i +\frac{1}{15}\sum_{k=1}^{15} (P_B^{k_i}+P_{\text{curt}}^{k_i})\big)^-h.
\end{align}
The \textbf{loss of consumer gain (LCG)} due to location is denoted as
\begin{equation}
 \text{LCG}_j = C_{\text{inv}}^j - C_{\text{wic}}^j.
\end{equation}
\textbf{Prosumer energy curtailed:} Nodes with greater voltage fluctuations result in a more restricted range of inverter operation, thus leading to greater energy curtailment. The \textbf{total curtailed energy (TCE)} for inverter at node $j$ is denoted as
\begin{equation}
 \text{TCE}_j = \sum_k P_{\text{curt}}^{(j,k)}.
\end{equation}
\textbf{Prosumer voltage correction:} 
%Consider uncontrolled voltage at node $j$ and time $k$ as \textcolor{red}{$U_{j,k}^{\text{uncontrol}}$} and the voltage after inverter control as \textcolor{red}{$U_{j,k}^{\text{correct}}$}. 
The \textbf{Voltage correction index (VCI)}
\textcolor{black}{consists of four metrics that shows 
    \begin{itemize}
        \item cumulative instances where voltage is beyond $V_{\max}$,
        \item cumulative instances where voltage is beyond $1+\Delta_{\text{perm}}$,
        \item cumulative instances where voltage is below $1-\Delta_{\text{perm}}$,
        \item cumulative instances where voltage is below $V_{\min}$.
    \end{itemize}
VCI}
for voltage time series $U$ is given as
\begin{equation}\begin{split}
 \text{VCI}(U) = \Big[\sum_k \mathbf{1}(U > U_{\max}), ~~ \sum_k \mathbf{1}(U > 1 + \Delta_{\text{perm}}), \\ \sum_k \mathbf{1}(U < 1 - \Delta_{\text{perm}}), \sum_k \mathbf{1}(U < U_{\min}) \Big],
 \end{split}
\end{equation}
where $\mathbf{1}(\text{condition})$ returns 1 if the condition is true, else it returns 0.
VCI provides the number of samples where the voltage level is outside the permissible level. The cumulative voltage correction can be identified by adding all voltage deviation outside $[1 - \Delta_{\text{perm}}, 1 + \Delta_{\text{perm}}]$ denoted as
\textbf{cumulative voltage correction (CVC).} CVC for voltage U is given as
\begin{equation}\begin{split}
 \text{CVC}(U) = \sum_k \mathbf{1}(U > 1 + \Delta_{\text{perm}})(U - 1 - \Delta_{\text{perm}}) ~+ \\ \sum_k \mathbf{1}(U < 1 - \Delta_{\text{perm}})(1 - \Delta_{\text{perm}} - U).
 \end{split}
\end{equation}
Next, we use the performance indices proposed to quantify the locational impact on prosumer energy optimization in presence of inverter rules based on voltage measurement at PCC.

\pagebreak
\section{Numerical results}
\label{section5}
Three numerical simulations are presented in this section. 
In Section \ref{subsec51}, energy optimization presented in Section \ref{section2} is implemented. 
The second numerical experiment in Section \ref{subsec52} combines the energy optimization with inverter P(U) and Q(U) rules for a prosumer. The effect of two controls PRC and ANRC with varying inverter sizes on performance indices are presented. Based on historical voltage variation, this exercise can be used to select the best-suited inverter.
The DN locational dependency of prosumer inverter control is explored in Section \ref{subsec53}.
A stylized DN with 3 identical consumers at different positions in the distribution feeder is analyzed for (a) their cost of energy consumption, (b) voltage correction ability, and (c) renewable energy curtailment.
Performance indices presented in Section \ref{section4} are used to compare different cases.

\subsection{Energy optimization of storage and load flexibility}
% \textcolor{red}{do we need this section.. with only energy optimization.. has no meaning about location or effect or inverters, unless we are comparing.. even then there is no need for thsi remove it.}
\label{subsec51}
\begin{figure}[!htbp]
	\center
	\includegraphics[width=0.8\linewidth]{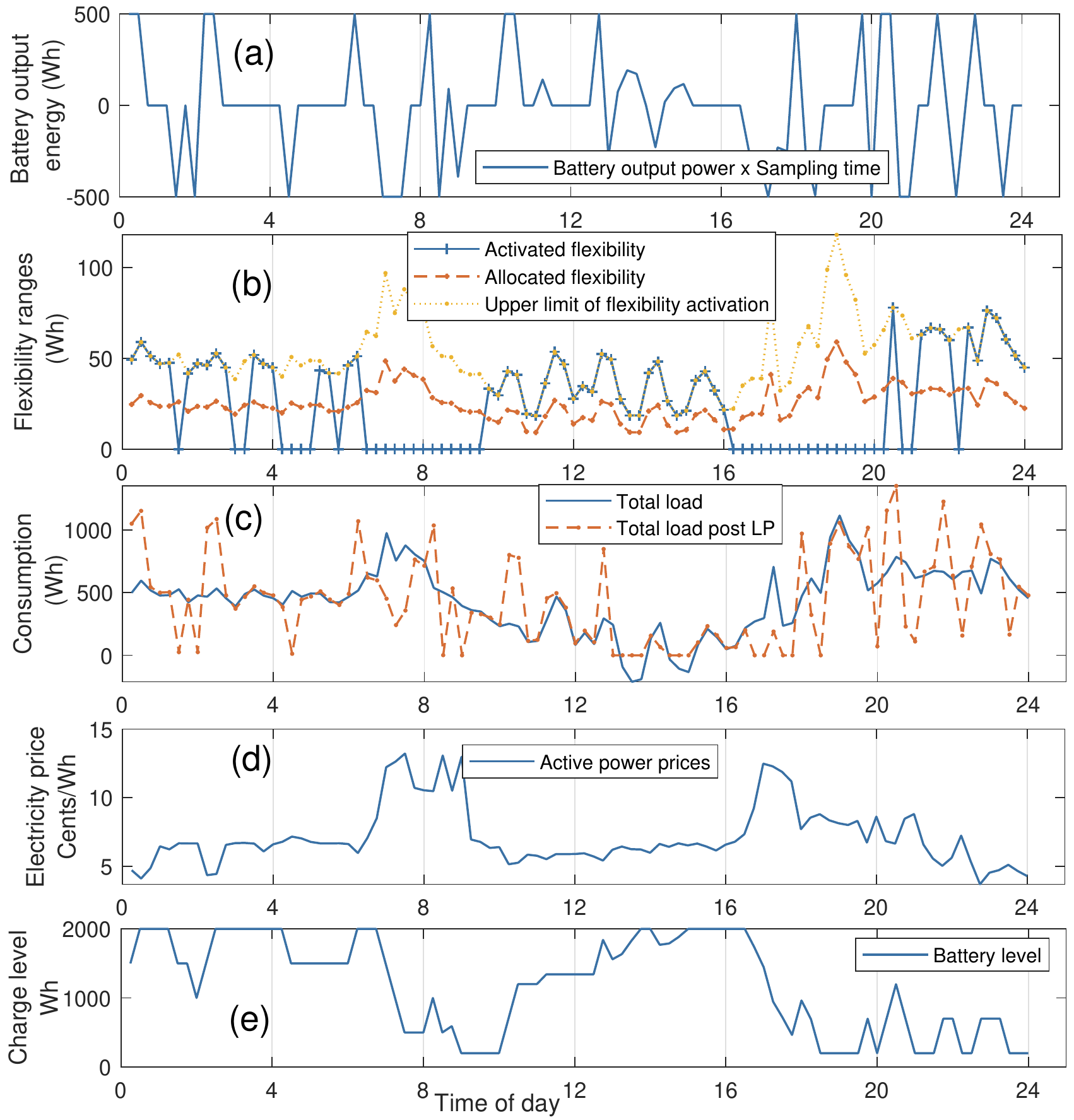}
	\vspace{-4pt}
	\caption{Numerical validation for activating load flexibility and energy storage arbitrage; 5\% load flexibility, $p_s/p_b = 0.5$}
	\label{fig:numericalres1}
\end{figure}
The results for energy management with 5\% load flexibility, 2 kWh 1C-1C battery\footnote{1C-1C battery takes 1 hour to charge fully and 1 hour to discharge fully from completely discharged and charged state respectively.}, 1.8 kWp solar is presented in Fig.~\ref{fig:numericalres1}. Electricity price data is used from \textcolor{black}{New York Independent System Operator or NYISO} wholesale real-time electricity price.
% \textcoor{blue}{Note that real-time wholesale electricity price is considered as it is time varying} 
The load flexibility bounds are from 0 to twice the available flexibility. The deadline constraint ensures that the energy consumption of the flexible resources is respected.
Fig.~\ref{fig:numericalres1} (a) shows the battery ramp rate, (b) the operation and bounds of load flexibility, (c) the total billable load, (d) buying price of electricity, and (e) the battery charge level.
The flexible load output shows that it is activated in a way to avoid morning and evening peaks in electricity prices, thus reducing the cost of consumption. Since $\kappa = 0.5$, self-consumption is prioritized which is evident from Fig.~\ref{fig:numericalres1} (c) where the net load after energy optimization saturates at zero, therefore, avoiding injection of active power in DN.
% \vspace{-6pt}
\begin{figure}[!htbp]
	\center
	\includegraphics[width=4.4in]{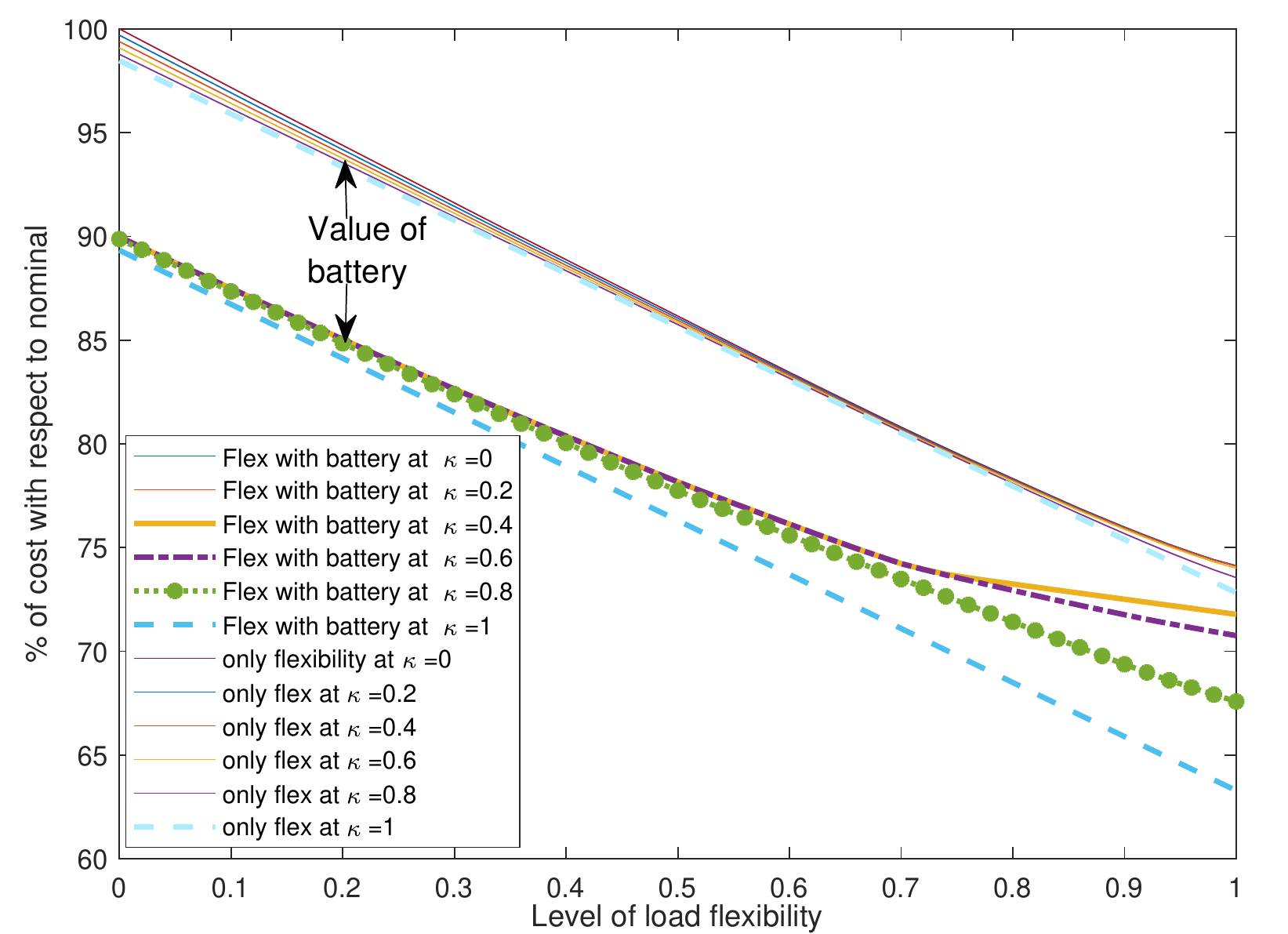}
	\vspace{-4pt}
	\caption{Value of storage, flexibility with varying load flexibility levels and $\kappa$ which is the ratio of selling price and buying price of electricity}
	\label{fig:numericalKappa}
\end{figure}

Fig.~\ref{fig:numericalKappa} shows the decrease in the cost of consumption with load flexibility without and with battery. Prosumers can reduce their cost of consumption by more than 25\% by selecting a consumption profile without reducing energy consumption.
Observe that load flexibility control is immune to $\kappa=p_s/p_b$. However, for batteries, the "value of battery" is higher for a lower value of $\kappa$. 

\subsection{Inverter controls with(out) energy optimization}
\label{subsec52}
\begin{figure}[!htbp]
	\center
	\includegraphics[width=4.4in]{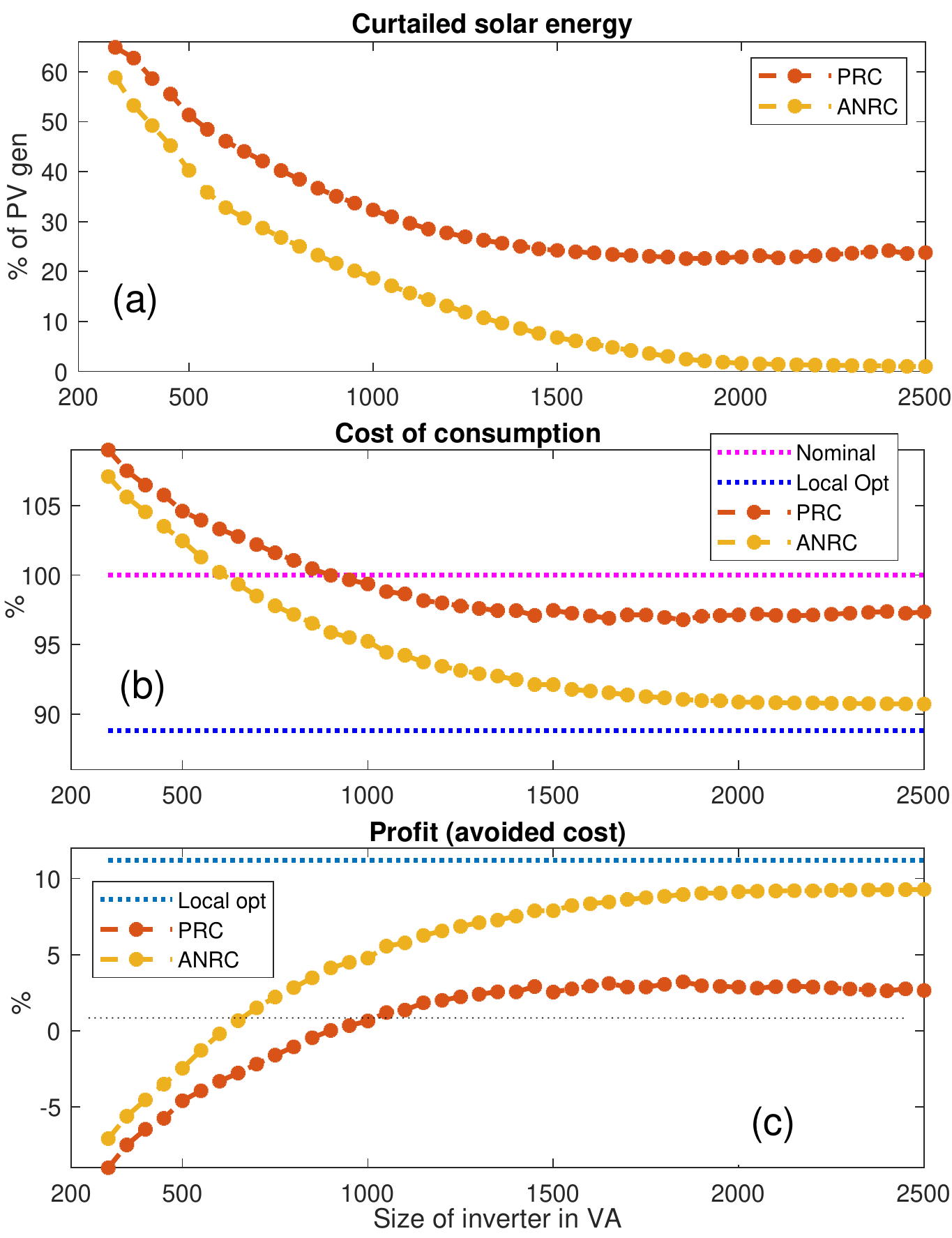}
	\vspace{-4pt}
	\caption{Inverter size impact on DG curtailed and consumer bill with $\kappa= 1$}
	\label{fig:numericalsimres2}
\end{figure}

Residential bidirectional inverters interfacing solar PV and battery
are expensive components. Selecting the size of the inverter is crucial in limiting the solar curtailed energy. In this numerical simulation, we perform a sensitivity analysis using (a) percentage solar energy, (b) percentage cost of consumption with respect to nominal load profile, and (c) percentage of profit with respect to only energy optimization without voltage control. Many DSOs motivate prosumers to self-consume their PV generation, this is ensured by setting $p_s<<p_b$. For this experiment, we assume $p_s = 0$.
The size of the solar panel is 2kWp and the battery capacity is 2 kWh (0.5C-0.5C, $\eta_{ch}=\eta_{dis} = 0.95$) requires $\approx$1.25 kW inverter for PRC inverter control and close to $\approx$2 kW for ANRC inverter control. 
Note from Fig.~\ref{fig:numericalsimres2} that the profit with ANRC \textcolor{black}{approaches} to energy optimization profit for a large-sized inverter.
Fig.~\ref{fig:numericalsimres2} also shows that for PV and battery, sharing an inverter will lead to a reduction in the required size of inverters compared to if individual inverters are used for PV and battery. For the oversized inverter, the profit with ANRC reaches 82.4\% and PRC reaches 25\% compared to only energy optimization.

Clearly, PRC-based inverter control will drastically reduce the energy optimization opportunities for a prosumer. Thus, it is anticipated that more volatile nodes at the end of a radial distribution feeder may have to pay more for energy compared to the same amount of energy consumed close to the substation, causing the locational disparity. In Section \ref{subsec53} we quantify this locational disparity.
\begin{center}
\begin{table}[!ht]
\centering
\small
	\caption {Line parameters for the network considered} 
	\label{tab:lineparameters}
	\begin{center}
		\begin{tabular}{|c|c|c|}
			\hline
			branch from \& to node & Resistance ($\Omega$) & \textcolor{black}{Reactance} ($\Omega$)\\
			\hline
			node 1 to 2 & 0.0922 & 0.0470\\
			node 2 to 3 & 0.1844 & 0.0940\\
			node 3 to 4 & 0.3660 & 0.1864\\
 \hline
		\end{tabular}
		\hfill\
	\end{center}
\end{table}
% \addtocounter{footnote}{-2}
\end{center}

\subsection{Effect of prosumer location in radial DN}
\label{subsec53}
\begin{figure}[!htbp]
	\center
	\includegraphics[width=0.9\linewidth]{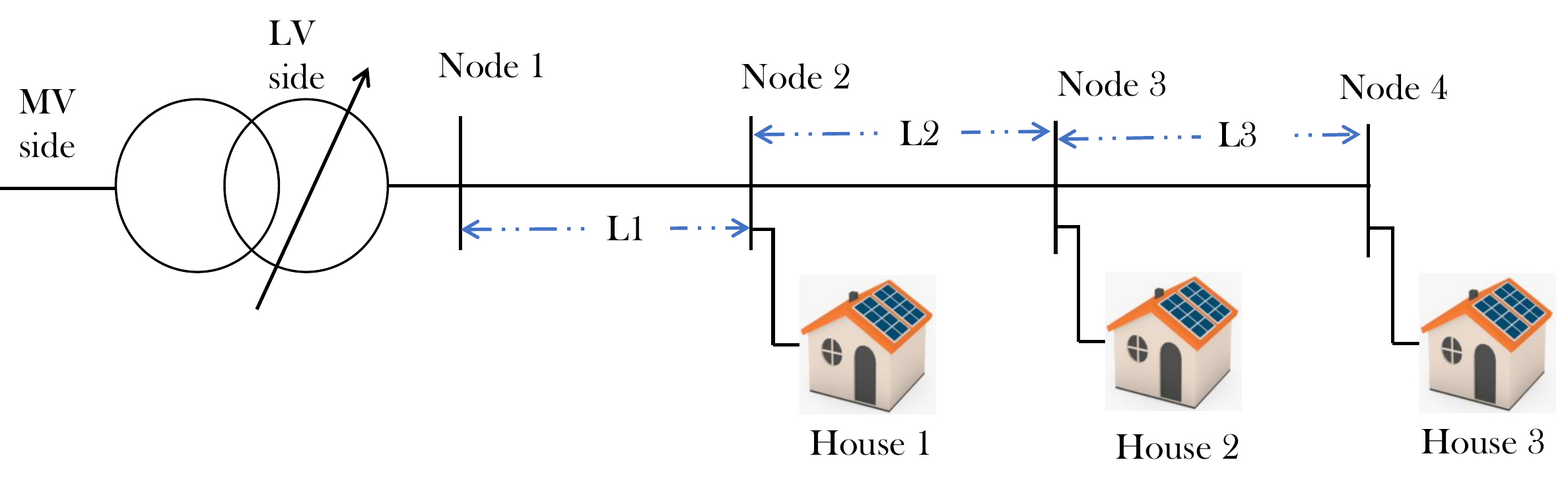}
	\vspace{-4pt}
	\caption{Stylized 4 bus radial DN with identical prosumers connected at bus 2,3,4}
	\label{fig:sysfair}
\end{figure}

DN feeders are predominantly radial. These feeders are exposed to different levels of voltage fluctuations depending on their location on the feeder. For nodes close to the feeder, the fluctuation is small compared to node buses away from the feeder. We consider 4 bus simple distribution feeders. The network diagram is shown in Fig.~\ref{fig:sysfair}. 
The line parameters of the 4 bus network considered in this work \textcolor{black}{are} listed in Table \ref{tab:lineparameters}.
% The branch between node 1 and 2 has r = 0.0922 $\Omega$,	x = 0.0470 $\Omega$. The branch between node 2 and 3 has	
% r = 0.1844 $\Omega$, x = 0.0940 $\Omega$. The branch between node 3 and 4 has r = 0.3660 $\Omega$, x = 0.1864 $\Omega$. \textcolor{red}{prof asks to insert table.}
The minimum and maximum voltage are assumed to be $V_{\min} = 0.92$ and $V_{\max} = 1.08$.
The permissible voltage level $\Delta_{\text{perm}} = 0.04$.
For this numerical example, the inverter is oversized so as to not curtail energy due to the capacity limitation of the inverter. The PV size at the prosumer end is 2.5 KWp and a 1 kW 0.5C-0.5C battery is connected to the 3 kVA bidirectional inverter.

\begin{figure}[!htbp]
	\center
	\includegraphics[width=0.8\linewidth]{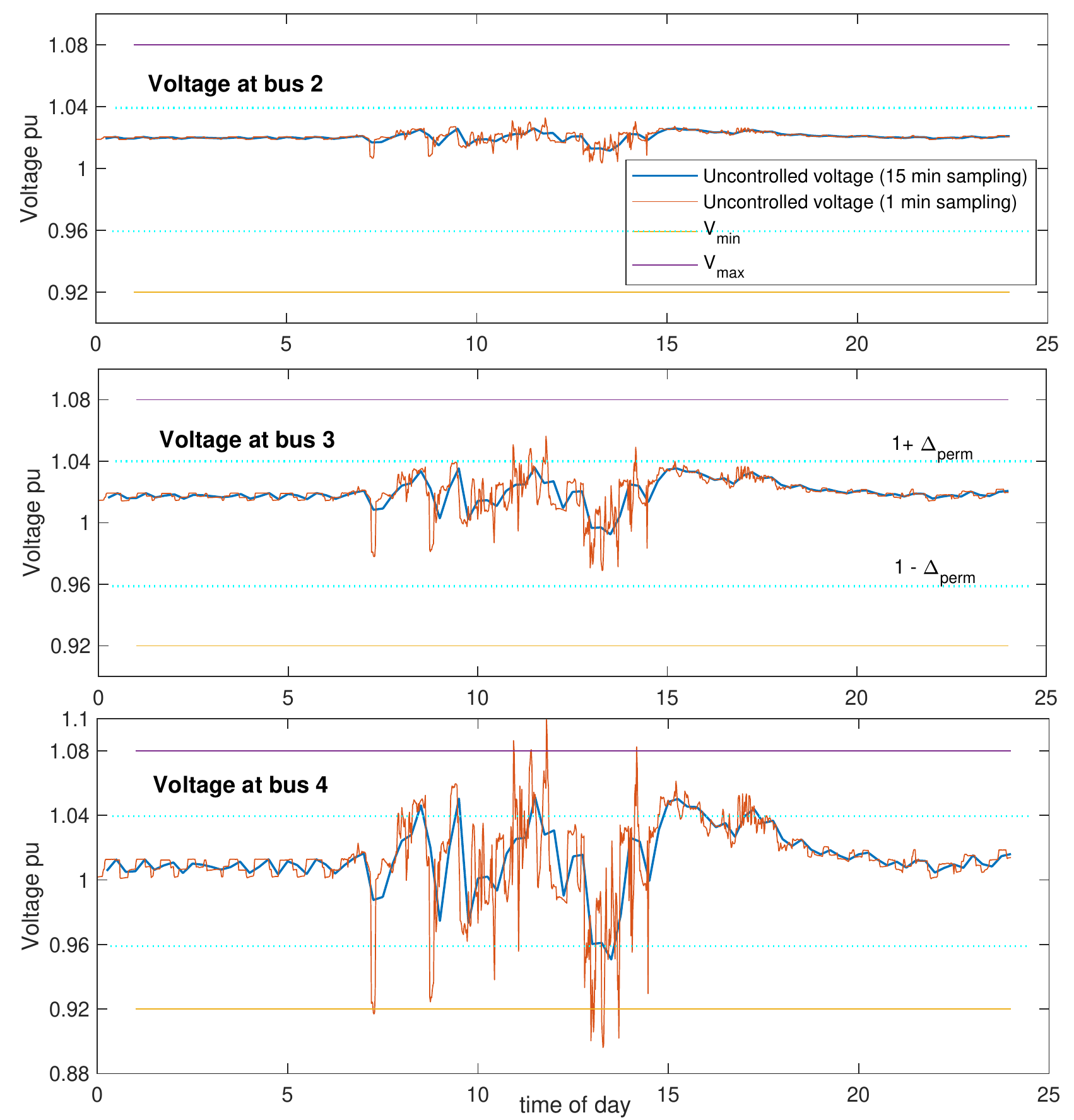}
	\vspace{-4pt}
	\caption{Nodal voltage in 3 bus radial distribution feeder}
	\label{fig:feedervoltage}
\end{figure}

The loads at nodes 2, 3, and 4 consist of a residential prosumer with identical load and solar generation.
Energy optimization with/without voltage-based inverter control is implemented at one node at a time.
The nodal voltage in absence of inverter control is shown in Fig.~\ref{fig:feedervoltage}.
The voltage at node 2 shows minimal fluctuations compared to node 4.
The following controls are evaluated and listed in Table~\ref{tab:disparity}:
\begin{itemize}
 \item Prosumer with no energy optimization,
 \item Energy optimization without inverter rules,
 \item Energy optimization with P(U) and Q(U) for ANRC,
 \item Energy optimization with P(U) and Q(U) for PRC,
 \item Energy optimization with hybrid policy: P control using ANRC and Q based on PRC.
\end{itemize}
% \vspace{-8pt}

% \subsection{Fairness}

\begin{table*}[!htbp]
\caption{Effect of inverter control on performance indices}
\centering
\resizebox{\linewidth}{!}{% use resizebox with textwidth
\begin{tabular}{|c|c|c|c|c| c|c|c|c|c |c|c| c|c|}
\hline
 \multicolumn{2}{|c|}{ Mode} & \multicolumn{3}{|c|}{ {Nominal case}} & \multicolumn{9}{|c|}{ {\textbf{With inverter control}}} \\ \hline
 & &\multicolumn{3}{|c|}{(with energy opt.)} & \multicolumn{3}{|c|}{\textbf{PRC}} & \multicolumn{3}{|c|}{\textbf{ANRC}} & \multicolumn{3}{|c|}{\textbf{Hybrid}} \\ \hline
\multicolumn{2}{|c|}{Node id} & 2 &3& 4 & 2 &3 &4 & 2 &3 &4 &2 &3 & 4 \\ \hline
\multirow{4}{*}{VCI} & $\sum \mathbf{1}(V > V_{\max})$ & 0	&0&	4&	 0&	0&	0&	0&	0&	0 & 0&	0 & 0 \\ 
&$\sum \mathbf{1}(V > 1 + \Delta_{\text{perm}})$&	0&	3&	86&				0&	3&	0&	0&	3	&82 & 0& 3&71\\ 
& $\sum \mathbf{1}(V < 1 - \Delta_{\text{perm}})$ & 0	&0	&39	&			0&	0&	48&	0&	0&	35 & 0&0&35\\ 
& $\sum \mathbf{1}(V <V_{\min})$	&0&	0&		1&		0&	0&	0&	0&	0	&1 & 0&0&0\\ \hline
\multicolumn{2}{|c|}{CVC}	&	0&	0.006&	1.71&		0&	0.006 &	0.637&	0&	0.006&	1.345 &0&0.006& 0.329\\ \hline
\multicolumn{2}{|c|}{Consumption cost (\euro ~cents)} &	37.78&37.78&37.78&			37.78 &	38.14 &	54.16 &	37.78 &	37.78 &	38.27 &37.78&37.78& 38.27\\ \hline
\multicolumn{2}{|c|}{LCG (\%)}	&	-&	-&	-&		0\%&	0.95\%&	43.35\%&	0\%&	0\%&1.3\% & 0\%&	0\%&1.3\%\\ \hline
\multicolumn{2}{|c|}{TCE (kWh)}	&	0&	0&		0&	0&	0.102 &	2.275&	0&	0&	0.188 & 	0&	0&0.188\\ \hline
 
\end{tabular}
}
 \label{tab:disparity}
\end{table*}

Table~\ref{tab:disparity} summarizes the results.
The nodal voltages are calculated using power flows performed using MATPOWER \cite{zimmerman2010matpower}.
Since the nodal voltage control is performed in an open loop, therefore, we observe the voltage deterioration for some time instances due to energy optimization, although the correction is performed for cases where voltage is outside its bound.
The prosumer connected at node 4 has LCG increases up to 43.35\%, refer to Table~\ref{tab:disparity}, causing increased consumption cost. 
% This implies consumer at node 4 with same consumption profile has to pay 43.35\% more compared to consumer at node 2. 
For ANRC, the curtailed energy is significantly reduced compared to PRC inverter control. ANRC does correct the node voltage, however, the correction is lower compared to PRC. 
ANRC is more favorable for consumer optimization as LCG is reduced from 43.35\% for PRC to a mere 1.3\%. 
Using ANRC will be fairer for prosumers located away from the feeder.
Note that for only energy optimization at node 4, the voltage profile deteriorates the most. This implies prosumer energy optimization needs to consider the network state into account.

For the prosumer located at node 2, the local voltage is always within bounds due to its proximity to the DN substation. Therefore, the prosumer does not have to bear any loss in energy optimization due to inverter rules and also does not have to provide any Q. This is also observed in \cite{demirok2011local}. One could observe that a prosumer connected at node 4 is at a disadvantage due to its location in the network and is obliged to supply Q and limit energy optimization gains without receiving any additional benefits compared to a prosumer at node 2.

\begin{figure}[!htbp]
	\center
	\includegraphics[width=0.9\linewidth]{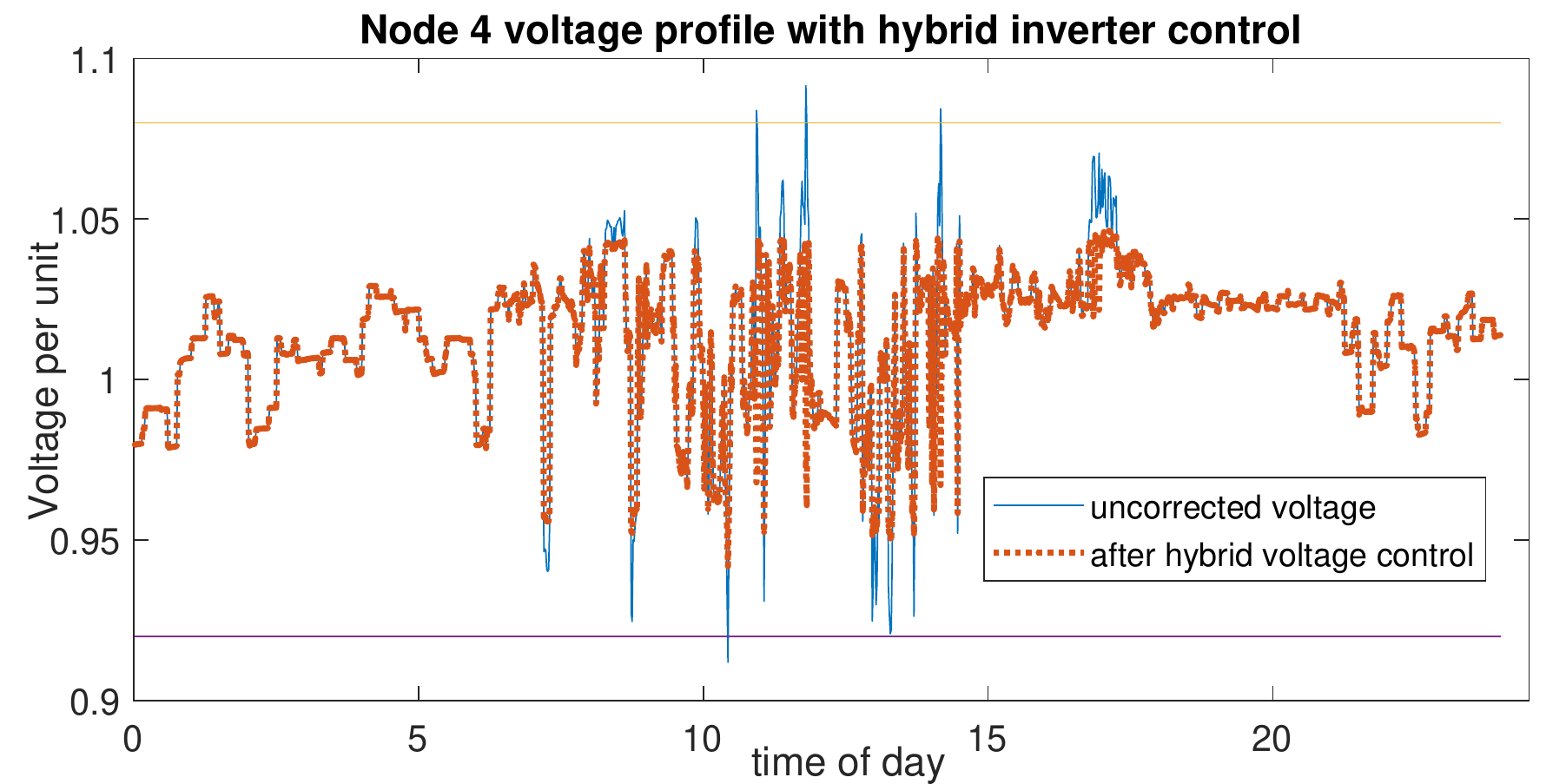}
	\vspace{-4pt}
	\caption{Corrected voltage at node 4 for hybrid controller}
	\label{fig:hybrid}
\end{figure}

The hybrid inverter control uses ANRC for active power control and uses PRC for reactive power control. 
% Depending on network R/X ratio, the Q compensation can correct voltage issues more efficiently. 
For our network example, R/X = 2. The hybrid inverter control outperforms only PRC in terms of CVC. For the hybrid model, CVC at node 4 is 0.3286 compared to 0.637 for PRC. 
Fig.~\ref{fig:hybrid} compares the voltages with only prosumer energy optimization and with energy optimization plus inverter voltage control.
% \begin{figure}[!htbp]
% 	\center
% 	\includegraphics[width=3.4in]{vcomp.eps}
% 	\vspace{-10pt}
% 	\caption{Comparing corrected voltage at node 4}
% 	\label{fig:comparisonvoltage}
% \end{figure}
% \vspace{-8pt}

\begin{figure}[!htbp]
	\center
	\includegraphics[width=0.8\linewidth]{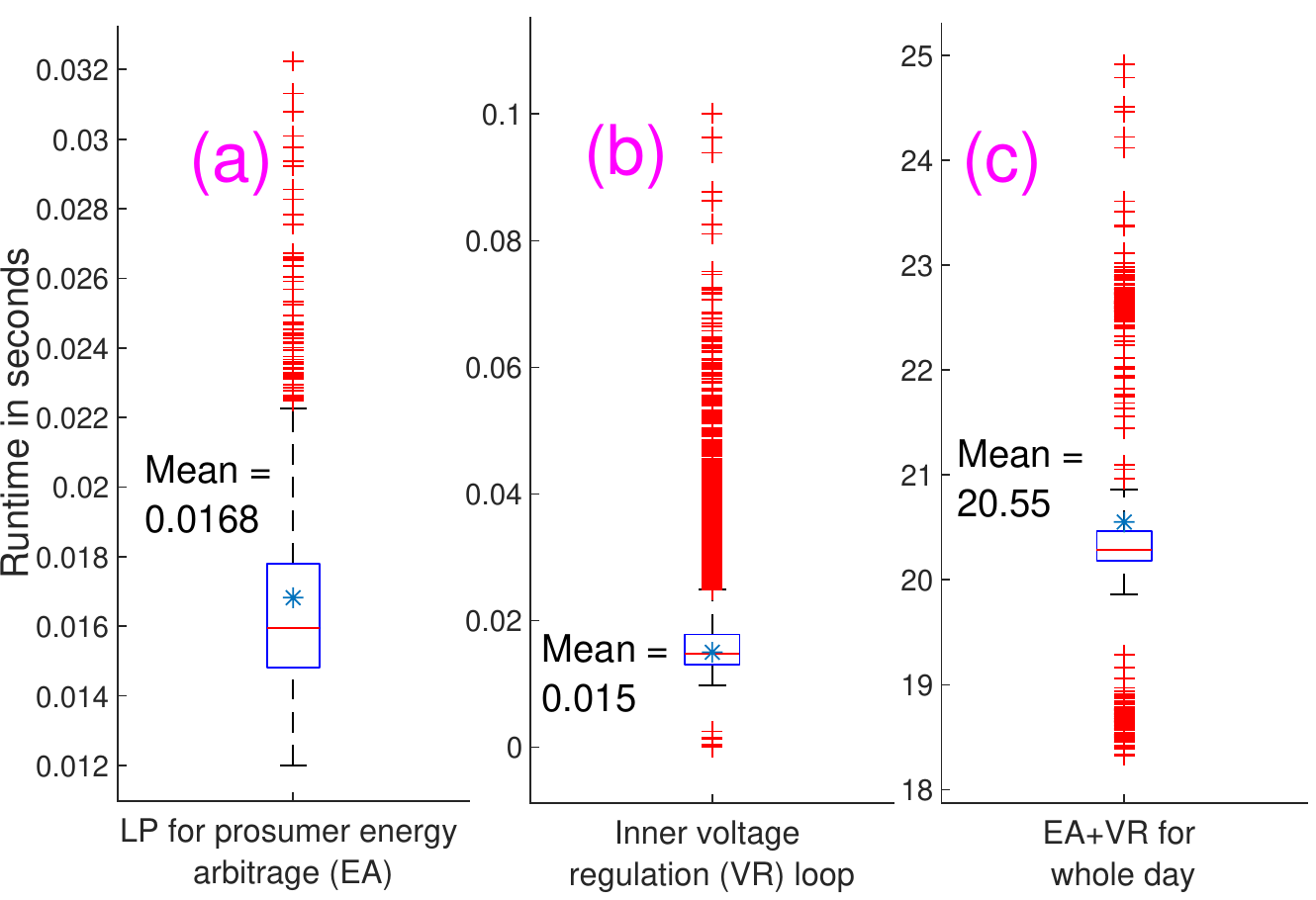}
	\vspace{-4pt}
	\caption{\textcolor{black}{Computational time taken for energy arbitrage and voltage regulation for 10000 Monte Carlo runs. (a) energy arbitrage based on LP for one entire day, (b) Inner loop runtime with 1 iteration of LP for energy arbitrage in receding horizon, (c) Total runtime for local energy optimization with voltage regulation for the entire day. The blue star shows the mean value.}}
	\label{fig:runtime}
\end{figure}

\textcolor{black}{
For the simulations, the computation time is calculated using 10000 Monte Carlo simulation runs. 
Simulations are performed on HP Intel(R) Core(TM) i7 CPU, 1.90GHz, 32 Gb RAM personal computer on Matlab 2021a.
The runtime evaluations for energy arbitrage based on LP (see Fig. \ref{fig:runtime}(a)) is done separately from run times for voltage regulation in the inner loop (see Fig. \ref{fig:runtime}(b)). Note that the mean runtime for energy arbitrage  is higher than the meantime of the inner loop with 1 iteration of energy arbitrage every 15 minutes because of the shrinking time horizon and because arbitrage is performed only once for the next 15 minutes. Key observations for runtime are
\begin{itemize}
    \item LP-based energy arbitrage takes on average 0.0168 seconds for running 96 times for the entire day. 
    \item The inner loop which performs voltage regulation includes (a) time taken to perform Power flows that depend on the network size, (b) one iteration of energy arbitrage every 15 minutes, and (c) voltage regulation performed every minute. For the test network, 1440 power flows are performed for 1 day, each iteration on average takes 0.015 seconds.
    \item The total runtime for energy arbitrage, power flows, and voltage regulation takes on an average of 20.55 seconds. Monte Carlo simulations for 1000 days were performed.
\end{itemize}
 Thus, the proposed voltage regulation with local energy optimization is computationally efficient.
}

\pagebreak
\section{Conclusion and discussion}
\label{section6}
We propose an LP formulation to control energy storage and load flexibility while considering electricity price variation, load profile, and renewable generation over a time horizon. This optimization is performed at a slower timescale. The output of local energy optimization is taken as input to update inverter output based on grid rules for inverter output on the basis of local voltage measurement. The P(U) and Q(U) inverter control is translated into positive reinforcement control (PRC) and avoiding negative reinforcement control (ANRC). The inverter control is performed at a finer timescale. Algorithms are presented to consider these different time scales, inverter size, locally measured voltage, and myopic curtailment reduction. 

We quantify the prosumer "loss of consumer gain" due to their location. A stylized network-based analysis is performed for a radial distribution network. \emph{Numerical results for a DN feeder indicate the prosumer connected at the end of the feeder may have to pay 43\% more than prosumers close to the feeder}. DSOs and regulators should hence consider that prosumer location could cause a disparity in their cost of consumption.
Hybrid inverter control with P control using 
ANRC 
and Q control using PRC along with P-priority significantly reduces the disparity for prosumers located at more vulnerable nodes while significantly improving the nodal voltage. This control methodology can be utilized to minimize the locational disparity caused by voltage-based inverter control from a prosumer perspective.

Inverter sizing is crucial for ensuring the full use of local resources such as energy storage, PV generation, and network connection rules.
We observe through numerical results that bidirectional inverter architecture where residential renewable generation and battery share a single inverter could lead to a significant reduction in its size compared to dedicated solar and battery inverters.

The mapping of voltage into P, and Q feasible ranges can be utilized in active distribution network design while using these ranges for cost optimization. Future work will utilize a forecast of nodal voltages for real-time operation and analyze the impact of inverter control on feeder hosting capacity.
\textcolor{black}{Further assessment is required to optimally set and numerically verify convergence of the droop slopes for envelope generation while taking into account the location of a prosumer in the DN. Finally, we will compare the centralized dispatch of resources for voltage regulation with our proposed distributed framework developed in this work.}

\pagebreak

\section*{Acknowledgement}
We would like to thank David Brummund and Maik Staudt from Mitnetz Strom for providing detailed inverter operational rules in Eastern Germany.
% This work is supported by the H2020 EUniversal project, grant agreement ID: 864334 (\url{https://euniversal.eu/}) and Moonshot Catalisti projects (\url{https://catalisti.be/moonshot/}).

\bibliographystyle{IEEEtran}
\bibliography{reference}

% \onecolumn
\pagebreak

% \onecolumn
\appendix

\subsection{LP formulation for arbitrage with energy storage and price-based flexibility dispatch}
\label{appendix:lpmatrix}

In our earlier work \cite{hashmi2019optimal}, we \textcolor{black}{used} linear programming (LP) based formulation to solve the optimal energy storage arbitrage problem. This LP formulation uses epigraph-based minimization \cite{boyd2004convex}. 
We extend the LP formulation in \cite{hashmi2019optimal, hashmi:tel-02462786} for including control of flexible load under time-varying electricity price. The formulation in \cite{hashmi2019optimal} uses the geometry of the cost function to form 4 line segments over which the epigraph-based cost function is minimized. Since flexible load $y_i \geq 0$, the operation of flexible loads does not lead to adding more line segments to the geometry of the cost function in \cite{hashmi2019optimal, hashmi:tel-02462786}. The LP formulation for solving the arbitrage problem is given as $P_{LP}$. The objective function and associated linear constraints are given in \eqref{eq:lpform}.

\begin{subequations}
\label{eq:lpform}
\begin{equation}
 (P_{LP}) ~~\min \quad \{t_1 + t_2+...+t_N\},
\end{equation}
\text{subject to,} \vspace{-10pt}
\begin{gather}
 \text{Segment 1:~}\frac{p_b^i}{\eta_{ch}} x_i + p_b^i y_i- t_i \leq -z_i p_b^i, ~\forall~ i,\\
 \text{Segment 2:~} {p_s^i}{\eta_{dis}} x_i + p_s^i y_i - t_i \leq -z_i p_s^i, ~\forall~ i,\\
 \text{Segment 3:~} {p_b^i}{\eta_{dis}} x_i + p_b^i y_i- t_i \leq -z_i p_b^i, ~\forall~ i,\\
 \text{Segment 4:~} \frac{p_s^i}{\eta_{ch}} x_i + p_s^i y_i- t_i \leq -z_i p_s^i, ~\forall~ i,\\
 \text{Ramp constraint:~} x_i \in [X_{\min}, X_{\max}], ~\forall~ i,\\
 \text{Capacity constraint:~} \sum {x_i} \in [b_{\min}-b_0, b_{\max}-b_0],~ \forall~ i,\\
 \text{Flexibility ramp constraint:~} \sum {y_i} \in [y_{\min}^i, y_{\max}^i],~ \forall~ i,\\
 \text{Cumulative flexibility:} 
\sum {y_i} \in [K-\epsilon, K+\epsilon],\text{for~} i=N.
\end{gather}
\end{subequations}

The matrix format for the optimization problem $P_{LP}$ is denoted as 
minimize
${f}^T X$, subject to ${A}X\leq b$, and $X \in [lb, ub]$.
The dimension of $A$ is (6N+2)x3N, $b$ is (6N+2)x1, $X$ and $f$ are of size 3Nx1, and N denotes the number of samples in the horizon of optimization.

\begin{equation}
f\text{=}{\begin{bmatrix}
	0\\
	:\\
	0\\
	0\\
	:\\
	0\\
	1\\
	:\\
	1\\
	\end{bmatrix}},
X \text{=} {\begin{bmatrix}
	x_1\\
	:\\
	x_N\\
	y_1\\
	:\\
	y_N\\
	t_1\\
	:\\
	t_N\\
	\end{bmatrix}},
\label{mateqsame0}
\end{equation}

\begin{equation}
lb\text{=}
{\begin{bmatrix}
	X_{\min}\\
	X_{\min}\\
	:\\
	X_{\min}\\
	y_{\min}(1)\\
	y_{\min}(2)\\
	:\\
	y_{\min}(N)\\
	T_{\min}\\
	T_{\min}\\
	:\\
	T_{\min}\\
	\end{bmatrix}}
\leq
{\begin{bmatrix}
	x_1 \\
	x_2 \\
	:\\
	x_N \\
	y_1 \\
	y_2 \\
	:\\
	y_N \\
	t_1 \\
	t_2 \\
	:\\
	t_N \\
	\end{bmatrix}} \leq
ub\text{=}
{\begin{bmatrix}
	X_{\max}\\
	X_{\max}\\
	:\\
	X_{\max}\\
	y_{\max}(1)\\
	y_{\max}(2)\\
	:\\
	y_{\max}(N)\\
	T_{\max}\\
	T_{\max}\\
	:\\
	T_{\max}\\
	\end{bmatrix}}, ~
	 {b} = {\begin{bmatrix}
	-z_i p_b(1) \\
	: \\
	-z_i p_b(N) \\
	-z_i p_s(1) \\
	: \\
	-z_i p_s(N) \\
	-z_i p_s(1) \\
	: \\
	-z_i p_s(N) \\
	-z_i p_b(1) \\
	: \\
	-z_i p_b(N) \\
	b_{\max} - b_0\\
	:\\
	b_{\max} - b_0\\
	b_0- b_{\min}\\
	:\\
	b_0- b_{\min}\\
	K+\epsilon\\
	-K + \epsilon
	\end{bmatrix}}.
\label{mateqsame}
\end{equation}
where $T_{\min}$ and $T_{\max}$ are bounds on $t_i$. Since these bounds are not known to us, we choose $T_{\min}$ to be negative with a large magnitude and $T_{\max}$ to be positive with a large magnitude.

{ \small{
\begin{equation}
{A} = { 
\renewcommand\arraystretch{1.3}
\mleft[ \begin{array}{c c c c c|c c c c c|c c c c c}
	\frac{p_b(1)}{\eta_{ch}} & 0 & 0 &... & 0 & p_b(1) & 0 & 0 &... & 0 & -1 & 0 & 0 &... & 0 \\
	0 & \frac{p_b(2)}{\eta_{ch}} & 0 &... & 0 & 0 & p_b(2) & 0 &... & 0 & 0 & -1 & 0 &... & 0 \\
	: & : & : &... & : & : & : & : &... & : & : & : & : &... & :\\
	0 & 0 & 0 &... & \frac{p_b(N)}{\eta_{ch}} & 0 & 0 & 0 &... & p_b(N) & 0 & 0 & 0 &... & -1 \\ \hline
	{p_s(1)}{\eta_{dis}} & 0 & 0 &... & 0 & p_s(1) & 0 & 0 &... & 0 & -1 & 0 & 0 &... & 0\\
	0 & {p_s(2)}{\eta_{dis}} & 0 &... & 0 & 0 & p_s(2) & 0 &... & 0 & 0 & -1 & 0 &... & 0 \\
	: & : & : &... & : & : & : & : &... & : & : & : & : &... & :\\
	0 & 0 & 0 &... & {p_s(N)}{\eta_{dis}} & 0 & 0 & 0 &... & p_s(N)& 0 & 0 & 0 &... & -1 \\ \hline
	\frac{p_s(1)}{\eta_{ch}} & 0 & 0 &... & 0 & p_s(1) & 0 & 0 &... & 0 & -1 & 0 & 0 &... & 0 \\
	0 & \frac{p_s(2)}{\eta_{ch}} & 0 &... & 0 & 0 & p_s(2) & 0 &... & 0 & 0 & -1 & 0 &... & 0 \\
	: & : & : &... & : & : & : & : &... & : & : & : & : &... & :\\
	0 & 0 & 0 &... & \frac{p_s(N)}{\eta_{ch}} & 0 & 0 & 0 &... & p_s(N) & 0 & 0 & 0 &... & -1\\ \hline
	{p_b(1)}{\eta_{dis}} & 0 & 0 &... & 0 & p_b(1) & 0 & 0 &... & 0& -1 & 0 & 0 &... & 0 \\
	0 & {p_b(2)}{\eta_{dis}} & 0 &... & 0 & 0 & p_b(2) & 0 &... & 0 & 0 & -1 & 0 &... & 0 \\
	: & : & : &... & : & : & : & : &... & : & : & : & : &... & :\\
	0 & 0 & 0 &... & {p_b(N)}{\eta_{dis}} & 0 & 0 & 0 &... & p_b(N)& 0 & 0 & 0 &... & -1 \\ \hline	
	1&0 & 0&...	&0 &0&0 & 0 &... & 0&0&0 & 0 &... & 0 \\
	1&1 & 0&...	&0 &0&0 & 0 &... & 0&0&0 & 0 &... & 0\\
	:&: & :&...	&: &:&: & : &... & :& : & : & : &... & : \\
	1&1 & 1&...	&1 &0&0 & 0 &... & 0&0&0 & 0 &... & 0 \\ \hline
	-1&0 & 0&...	&0 &0&0 & 0 &... & 0&0&0 & 0 &... & 0 \\
	-1&-1 & 0&...	&0 &0&0 & 0 &... & 0&0&0 & 0 &... & 0 \\
	:&: & :&...	&: &:&: & : &... &:& : & : & : &... & : \\
	-1&-1 & -1&...	&-1 &0&0 & 0 &... & 0&0&0 & 0 &... & 0 \\ \hline
	0&0 & 0&...	&0 &1&1 & 1 &... &1& 0 & 0 &0 &... & 0 \\
	0&0 & 0&...	&0 &-1&-1 & -1 &... &-1& 0 & 0 &0 &... & 0 
	\end{array} \mright]}, 
\end{equation} }
}

\subsection{LP formulation for arbitrage with energy storage and price-based flexibility dispatch}
\label{appendix:control_algo}
\textcolor{black}{We provide here, the details of the algorithm to determine curtailment, inverter active and reactive outputs over a faster timescale, discussed in Section \ref{subsec34}. Algorithm \ref{alg:control} used bounds $[ R^U_{P_{\min}}, R^U_{P_{\max}}]$ and $[ R^U_{Q_{\min}}, R^U_{Q_{\max}}]$ from the respective Tables \ref{tab:positiveLimits}, \ref{tab:negativeLimits}, and \ref{tab:hybridinverter} for different control policies (PRC, ANRC or Hybrid).
}
\begin{algorithm}
	\begin{algorithmic}[1]
	\State Use $f(x_i),r_i$ as global variables,
	%	 \Function{Calculate\_Inverter\_Output}{$\zeta_i, R_P^U, R_Q^U$}
	 \If{$\zeta_i \in [ R^U_{P_{\min}}, R^U_{P_{\max}}]$ and $Q_{\text{default}}\in [ R^U_{Q_{\min}}, R^U_{Q_{\max}}]$}
	 \State Set $P_{\text{inv}}^{k_i}= \zeta_i, Q_{\text{inv}}^{k_i}=0, P_{\text{curt}}^{k_i}=0$
	 \ElsIf{$\zeta_i \notin [ R^U_{P_{\min}}, R^U_{P_{\max}}]$ and $Q_{\text{default}}\in [ R^U_{Q_{\min}}, R^U_{Q_{\max}}]$}
	 \If{$\zeta_i < R^U_{P_{\min}}$}
	 \texttt{Active\_Point\_Minimize}($ R^U_{P_{\min}}$)
	 \ElsIf{$\zeta_i > R^U_{P_{\max}}$}
	 \texttt{Active\_Point\_Minimize}($ R^U_{P_{\max}}$)
	 \EndIf
	 \State Set $Q_{\text{inv}}^{k_i}=0$
	 \ElsIf{$\zeta_i \in [ R^U_{P_{\min}}, R^U_{P_{\max}}]$ \& $Q_{\text{default}}\notin [ R^U_{Q_{\min}}, R^U_{Q_{\max}}]$}
	 \If{$Q_{\text{default}} < R^U_{Q_{\min}}$}
	 ~~~ $Q_{\text{inv}}^{k_i} = R^U_{Q_{\min}}$
	 \ElsIf{$Q_{\text{default}} > R^U_{Q_{\max}}$}~~~ $Q_{\text{inv}}^{k_i} = R^U_{Q_{\max}}$
	 \EndIf
	 \State Set $P_{\text{inv}}^{k_i}=\zeta_i, P_{\text{curt}}^{k_i} = 0$
	 \ElsIf{$\zeta_i \notin [ R^U_{P_{\min}}, R^U_{P_{\max}}]$ \& $Q_{\text{default}}\notin [ R^U_{Q_{\min}}, R^U_{Q_{\max}}]$}
	 \If{$\zeta_i < R^U_{P_{\min}}$ and $Q_{\text{default}} < R^U_{Q_{\min}}$}
	 \State Execute \texttt{Active\_Point\_Minimize}($ R^U_{P_{\min}}$) to find $P_{\text{curt}}^{k_i}, P_{\text{inv}}^{k_i}$; Set $Q_{\text{inv}}^{k_i} = R^U_{Q_{\min}}$,
	 \ElsIf{$\zeta_i < R^U_{P_{\min}}$ and $Q_{\text{default}} > R^U_{Q_{\max}}$}
	 \State Execute \texttt{Active\_Point\_Minimize}($ R^U_{P_{\min}}$) to find $P_{\text{curt}}^{k_i}, P_{\text{inv}}^{k_i}$; Set $Q_{\text{inv}}^{k_i} = R^U_{Q_{\max}}$, 
	 \ElsIf{$\zeta_i > R^U_{P_{\max}}$ and $Q_{\text{default}} < R^U_{Q_{\min}}$}
	 \State Execute \texttt{Active\_Point\_Minimize}($ R^U_{P_{\max}}$) to find $P_{\text{curt}}^{k_i}, P_{\text{inv}}^{k_i}$; Set $Q_{\text{inv}}^{k_i} = R^U_{Q_{\min}}$,
	 \ElsIf{$\zeta_i > R^U_{P_{\max}}$ and $Q_{\text{default}} > R^U_{Q_{\max}}$}
	 \State Execute \texttt{Active\_Point\_Minimize}($ R^U_{P_{\max}}$) for $P_{\text{curt}}^{k_i}, P_{\text{inv}}^{k_i}$, and Set $Q_{\text{inv}}^{k_i} = R^U_{Q_{\max}}$,
	 \EndIf
	 \EndIf
 \State \Return $P_{\text{curt}}^{k_i}, P_{\text{inv}}^{k_i}, Q_{\text{inv}}^{k_i}$
 %\EndFunction
 % \vspace{10pt}
 
 \Function{Active\_Point\_Minimize}{$P^U_{trgt}$}
	 \State minimize $P_{\text{curt}}^{k_i}$ as described in \eqref{eq:findingcurtailment}, and Set $P_{\text{inv}}^{k_i} = P_B^{k_i} - r_i+P_{\text{curt}}^{k_i}$,
	 \If{optimization ($P_{LP}^{curt}$) does not provide a solution}
	 \If{$P^U_{trgt}>0$} $P_{\text{curt}}^{k_i} =r_{i}$ and $P_B^{k_i}=$ max charge,
	 \ElsIf{$P^U_{trgt}<0$} $P_{\text{curt}}^{k_i} =0$ and $P_B^{k_i}=$ max discharge,
	 \EndIf
	 \EndIf
% 	 \State If optimization ($P_{LP}^{curt}$) does not provide a feasible solution then set $P_{\text{curt}}^{k_i} =r_{k_i}$. If $R_P^U>0$ then 
% $P_B^{k_i}=$ max charge, else $P_B^{k_i}=$ max discharge
 \State \Return $P_{\text{curt}}^{k_i}, P_{\text{inv}}^{k_i}$
 \EndFunction
 
	\end{algorithmic}
\caption{\texttt{Local Inverter Output Control}}
	\label{alg:control}
\end{algorithm}

\pagebreak

    \begin{IEEEbiography}[{\includegraphics[width=.8in,height=1in,clip,keepaspectratio]{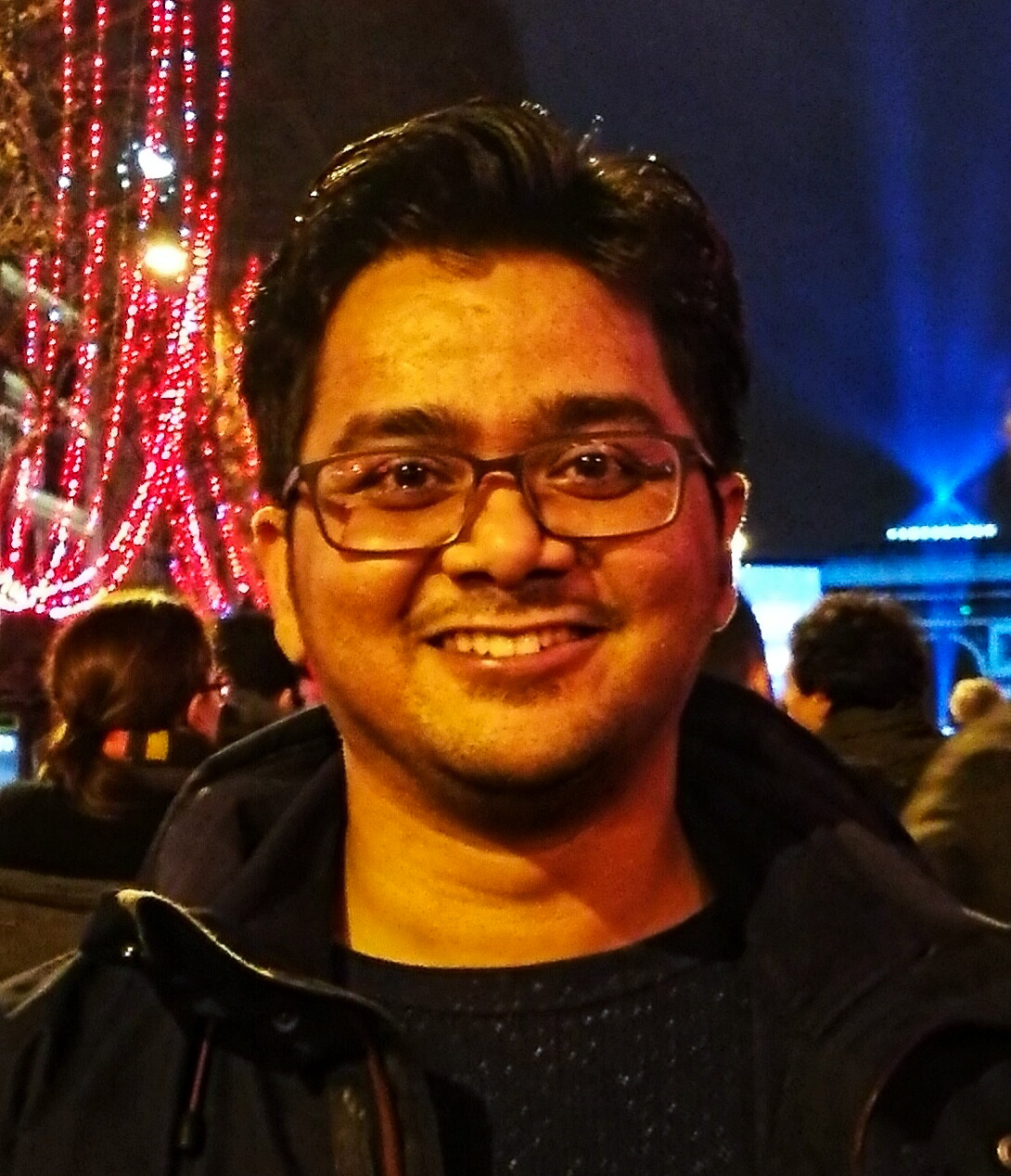}}]{Md Umar Hashmi}
    is a senior postdoctoral researcher at KU Leuven and EnergyVille in Belgium. He
       completed his PhD at \'Ecole Normale Sup\'erieure and INRIA, Paris France in December 2019. 
       He also worked for Eaton Corporation as a Controls Engineer and for Eirgrid in Dublin as Senior Engineer. %He holds a US Patent.
       He completed his master's and bachelor's degree from the Indian Institute of Technology Bombay in 2012 and Aligarh Muslim University in 2010, respectively. 
       His research interests include electrical power systems, smart grids, renewable integration, data analytics, control, and optimization algorithm development.
    \end{IEEEbiography}
  \vspace{-20pt}
    \begin{IEEEbiography}[{\includegraphics[width=.8in,height=1in,clip,keepaspectratio]{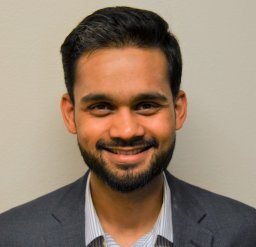}}]{Deepjyoti Deka}
       is a staff scientist in the Applied Mathematics and Plasma Physics group of the Theoretical Division at Los Alamos National Laboratory, where he was previously a postdoctoral research associate at the Center for Nonlinear Studies. His research interests include data analysis of power grid structure, operations and security, and optimization in social and physical networks. At LANL, Dr. Deka serves as a co-principal investigator for DOE projects on machine learning in distribution systems and in cyber-physical security. Before joining the laboratory he received the M.S. and Ph.D. degrees in electrical engineering from the University of Texas, Austin, TX, USA, in 2011 and 2015, respectively. He completed his undergraduate degree in electrical engineering from IIT Guwahati, India  in 2009 with an institute silver medal as the best outgoing student of the department. Dr. Deka is a senior member of IEEE.
    \end{IEEEbiography}
\vspace{-34pt}
    \begin{IEEEbiography}[{\includegraphics[width=.8in,height=1in,clip,keepaspectratio]{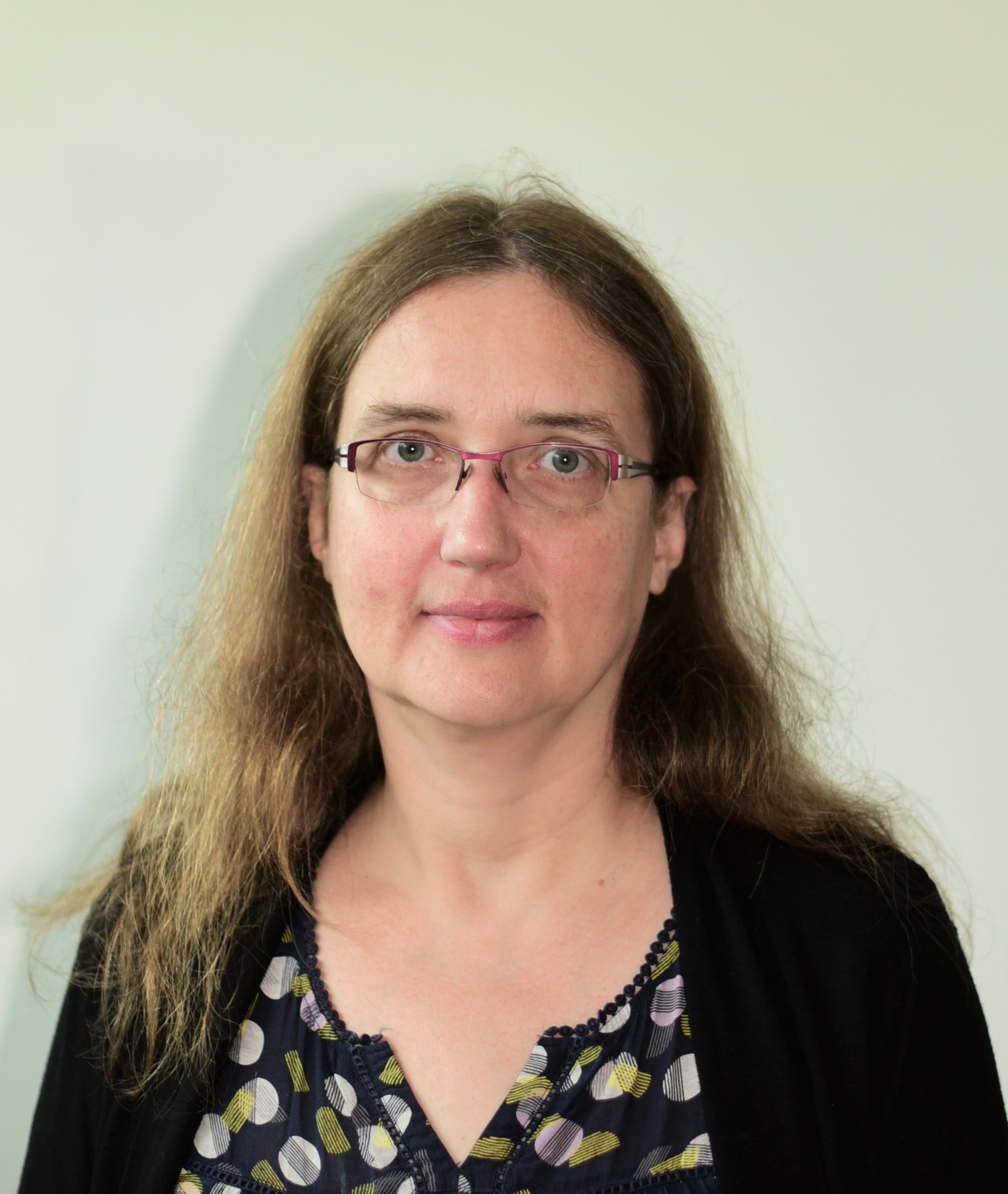}}]{Ana Bu\v{s}i\'c}
    is a Research Scientist at Inria Paris Research Centre and the Computer Science Department at Ecole Normale Sup\' erieure, PSL University, France. She received the M.S. degree in Applied Mathematics and the Ph.D. degree in Computer Science from the University of Versailles, France, in 2003 and 2007, respectively.  She was a Post-Doctoral Fellow at Inria Grenoble—Rh\^one-Alpes and at University Paris-Diderot-Paris 7. Her research interests include stochastic modeling, simulation, optimization, and reinforcement learning, with applications to communication networks and energy systems. She is a member of the Laboratory of Information, Networking, and Communication Sciences, a joint lab between Inria, Institut Mines-Télécom, UPMC Sorbonne Universities, Nokia Bell Labs, and SystemX. She received a Google Faculty Research Award in 2015. 
    \end{IEEEbiography}

    \begin{IEEEbiography}[{\includegraphics[width=.8in,height=1in,clip,keepaspectratio]{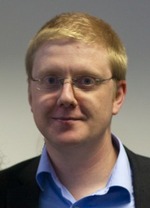}}]{Dirk Van Hertem}
     graduated as a M.Eng. in 2001 from the KHK, Geel, Belgium and as a M.Sc. in Electrical Engineering from the KU Leuven, Belgium in 2003. In 2009, he has obtained his PhD, also from the KU Leuven. In 2010, Dirk Van Hertem was a member of EPS group at the Royal Institute of Technology (KTH), in Stockholm. Since spring 2011 he is back at the University of Leuven where he is currently professor and member of the ELECTA division. His special fields of interest are decision support for grid operators, power system operation and control in systems with FACTS and HVDC and building the transmission system of the future, including offshore grids and the supergrid concept. The research activities of Prof. Van Hertem are all part of the EnergyVille research center, where he leads the Electrical Networks activities.  Dr. Van Hertem is an active member of both IEEE (PES and IAS) and Cigré. 
    \end{IEEEbiography}

\end{document}